\documentclass[conference]{IEEEtran}
\usepackage{cite}
\usepackage{amsmath,amssymb,amsfonts}
\usepackage{graphicx}
\usepackage{textcomp}
\usepackage{xcolor}
\usepackage[hyphens]{url}
\usepackage{physics}
\usepackage{algorithm2e}
\usepackage{comment}
\usepackage{float}
\usepackage{multirow}
\usepackage{svg}
\def\BibTeX{{\rm B\kern-.05em{\sc i\kern-.025em b}\kern-.08em
    T\kern-.1667em\lower.7ex\hbox{E}\kern-.125emX}}

\usepackage{subfig}
\graphicspath{ {images/} }
\usepackage{pgfplots, pgfplotstable}
\usepackage{filecontents}
\pgfplotsset{width = \linewidth,compat=1.9}
\usepackage[export]{adjustbox}
\usepackage{array}
\newcolumntype{P}[1]{>{\centering\arraybackslash}p{#1}}
\newcolumntype{M}[1]{>{\centering\arraybackslash}m{#1}}
\newcommand{\linebreakand}{%
  \end{@IEEEauthorhalign}
  \hfill\mbox{}\par
  \mbox{}\hfill\begin{@IEEEauthorhalign}
}

\title{Folding-Free ZNE: A Comprehensive Quantum
Zero-Noise Extrapolation Approach for Mitigating
Depolarizing and Decoherence Noise}

\author{\IEEEauthorblockN{%1\textsuperscript{st} 
Hrushikesh Pramod Patil}
\IEEEauthorblockA{\textit{NC State University}\\
Raleigh, NC, USA \\
hpatil2@ncsu.edu\\}
\and
\IEEEauthorblockN{%2\textsuperscript{nd} 
Peiyi Li}
\IEEEauthorblockA{\textit{NC State University}\\
Raleigh, NC, USA \\
pli11@ncsu.edu\\}
\and
\IEEEauthorblockN{%3\textsuperscript{rd} 
Ji Liu}
\IEEEauthorblockA{\textit{Argonne National Laboratory}\\
Lemont, IL, USA \\
ji.liu@anl.gov\\}
\and
\IEEEauthorblockN{%4\textsuperscript{th} 
Huiyang Zhou}
\IEEEauthorblockA{\textit{NC State University}\\
Raleigh, NC, USA \\
hzhou@ncsu.edu\\}

}
\begin{document}
\maketitle
\thispagestyle{plain}
\pagestyle{plain}

%%%%%% -- PAPER CONTENT STARTS-- %%%%%%%%

\begin{abstract}

Quantum computers in the NISQ era are prone to noise. A range of quantum error mitigation techniques has been proposed to address this issue. Zero-noise extrapolation (ZNE) stands out as a promising one. ZNE involves increasing the noise levels in a circuit and then using extrapolation to infer the zero noise case from the noisy results obtained. This paper presents a novel ZNE approach that does not require circuit folding or noise scaling to mitigate depolarizing and/or decoherence noise.

To mitigate depolarizing noise, we propose leveraging the extreme/infinite noisy case, which allows us to avoid circuit folding. Specifically, the circuit output with extreme noise becomes the maximally mixed state. We show that using circuit-reliability metrics, simple linear extrapolation can effectively mitigate depolarizing noise. With decoherence noise, different states decay into the all-zero state at a rate that depends on the number of excited states and time. Therefore, we propose a state- and latency-aware exponential extrapolation that does not involve folding or scaling. When dealing with a quantum system affected by both decoherence and depolarizing noise, we propose to use our two mitigation techniques in sequence: first applying decoherence error mitigation, followed by depolarizing error mitigation.

A common limitation of ZNE schemes is that if the circuit of interest suffers from high noise, scaling-up noise levels could not provide useful data for extrapolation. We propose using circuit-cut techniques to break a large quantum circuit into smaller sub-circuits to overcome this limitation. This way, the noise levels of the sub-circuits are lower than the original circuit, and ZNE can become more effective in mitigating their noises.

\end{abstract}

\section{Introduction}
Recent years have seen exciting advances in the development of quantum computers. However, current NISQ (Noisy Intermediate Scale Quantum) devices are highly susceptible to noise and have limited numbers of qubits with constrained connectivity. Imperfections in hardware and the interaction with the environment inevitably lead to computations riddled with errors and noise. Noise in quantum computers is the fundamental challenge facing the reliable execution of quantum algorithms. To deal with this problem, %While improvements in hardware will bring down the error rates, we cannot guarantee error-free computation in a quantum computer. 
quantum error correction (QEC) methods %like the Calderbank-Shor-Steane (CSS) codes, Stabilizer codes, etc., 
along with fault-tolerant quantum computation have been proposed. Although the physical error rates of quantum devices are approaching the critical thresholds, where QEC can achieve lower logical error rates than the physical one, the required number of qubits is overwhelming and beyond the capacity of current devices. 

Quantum error mitigation (QEM) can be viewed as an alternative or complementary to QEC. %Quantum Error Mitigation, as the name implies, mitigates the errors rather than completely correcting them. 
Generally, QEM methods involve multiple runs of a circuit with a few parameters, such as the circuit depth, pulse length, etc., being varied. The results of these multiple runs are post-processed to obtain a noise-mitigated result. %The benefit of QEM over QEC is that QEM can be practically applied to current NISQ devices to improve their noise resilience. %QEM scales for wider or deeper quantum circuits with improvement in error rates. 
Popular error mitigation methods include Zero Noise Extrapolation (ZNE)  \cite{li2017efficient, PhysRevLett.119.180509, giurgica2020digital}, Probabilistic Error Cancellation (PEC) \cite{PhysRevLett.119.180509}, Clifford Data Regression (CDR) \cite{Czarnik2021errormitigation}, %Virtual Distillation \cite{PhysRevX.11.041036_virt}, 
etc. Machine learning-based quantum error mitigation approaches like QRAFT (Quantum circuit Reversal for Attaining Full Truth) \cite{QRAFT} have also been proposed recently.

ZNE was first introduced in \cite{li2017efficient} and \cite{PhysRevLett.119.180509}. The basic idea is to deliberately run the circuit with different noise levels and use the measured results to extrapolate the noise-free result. 

This paper improves the efficacy of ZNE with the following novel designs. First, we propose to change the way how the extrapolation is performed such that we can leverage the extreme noise case. The quantum states under extreme noise are independent of the circuit structure but dependent on the noise source. With depolarizing noise, the quantum state under extreme noise becomes the maximally mixed state. In existing ZNE schemes, the extrapolation is based on noise levels or circuit folding factors. As a result, it is impractical to incorporate extreme noise cases, whose noise levels/folding factors are infinite. 

Second, we choose to use different extrapolation algorithms to mitigate different noises. We show that using circuit-reliability metrics, simple linear extrapolation is sufficient to mitigate depolarizing noise, while exponential extrapolation is needed for mitigating decoherence noise. For systems suffering from both depolarizing and decoherence noise, we propose a serial approach, which first mitigates the decoherence noise followed by depolarization error mitigation. In all these cases, there is no need for circuit folding for extrapolation.

Third, rather than scaling up the noise levels for extrapolation, we propose to scale down the noise levels using circuit cutting techniques, which break a large circuit into smaller sub-circuits and rely on classical processing to combine the execution results from the sub-circuits. This way, our scheme can tackle the long-standing problem of all existing ZNE schemes, which become ineffective for circuits with high noise levels as further scaling up the noise fails to produce meaningful data for extrapolation.

The remainder of the paper is organized as follows. Section II presents the background on quantum noises and discusses the related work. Section III motivates our approach. Section IV details our proposed schemes. Sections V and VI discuss the experimental methodology and the results, respectively. Section VII concludes.

\section{Background}
\subsection{Noise in Quantum Computers}

Noise in quantum computers can lead to errors during computation. Depolarizing, decoherence or thermal relaxation, state preparation, measurement, and cross-talk errors are the most common errors in a quantum computer. Among them, state preparation and measure (SPAM) errors can be suppressed with careful calibration and post-processing based on the calibration data (e.g., an inverse of the measurement error matrix). Cross-talk errors depend on the device topology and can be mitigated by scheduling two-qubit gates judiciously in a circuit \cite{crosstalk_scheduling}. In this work, we focus on depolarizing and decoherence noise. 
\begin{figure*}[h]
    \centering
    \includegraphics[width=0.33\linewidth]{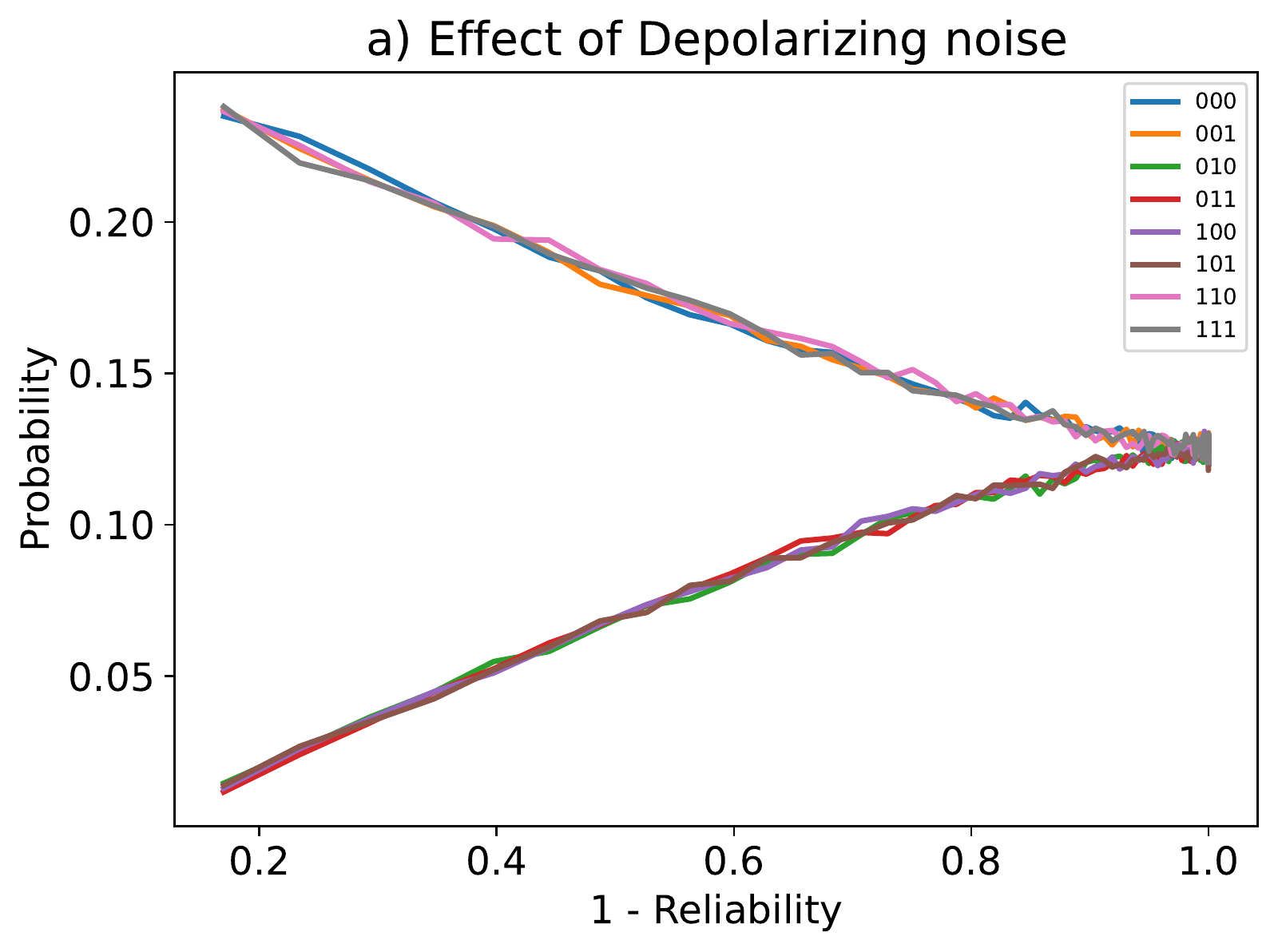}\hfill
    \includegraphics[width=0.34\linewidth]{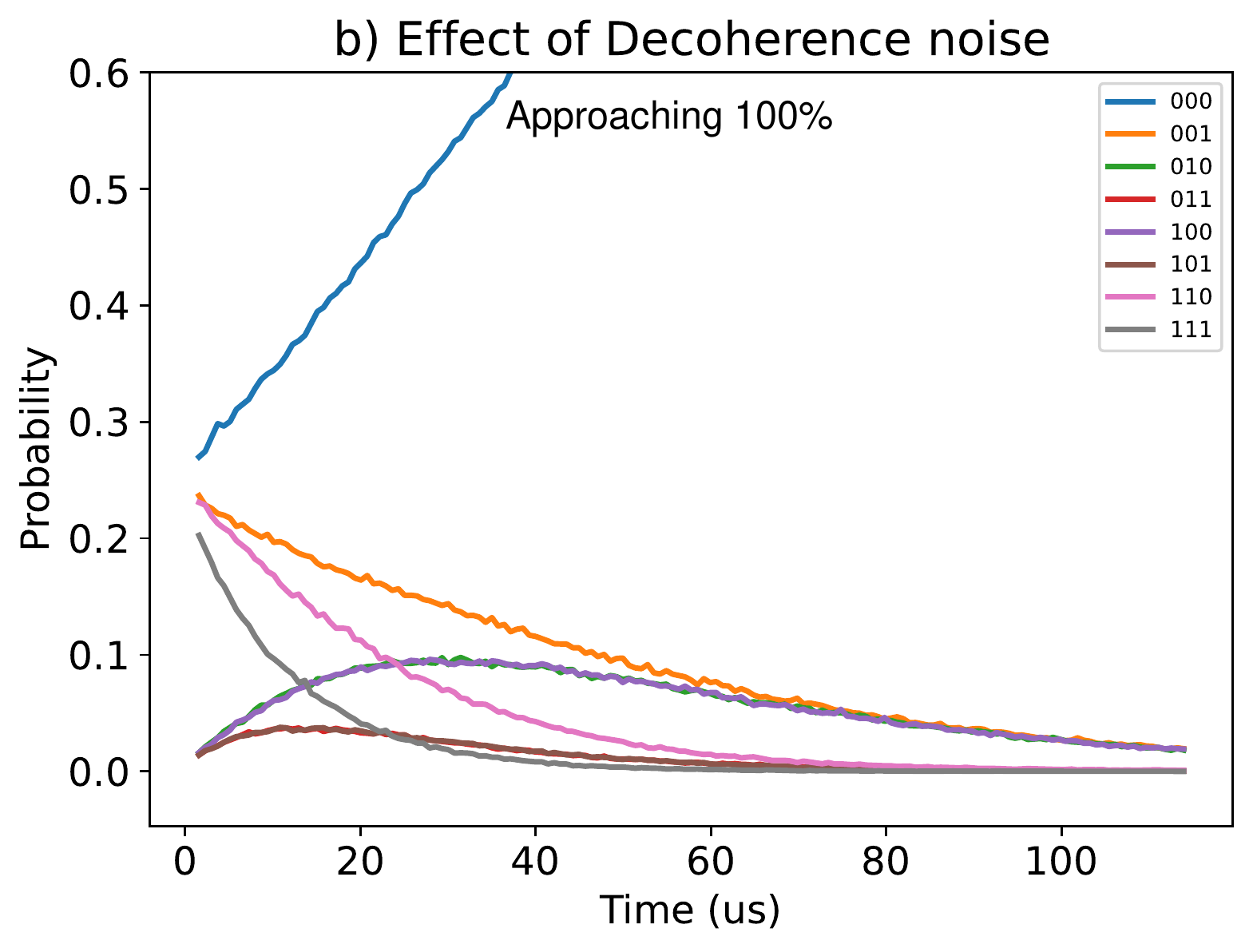}\hfill
    \includegraphics[width=0.33\linewidth]{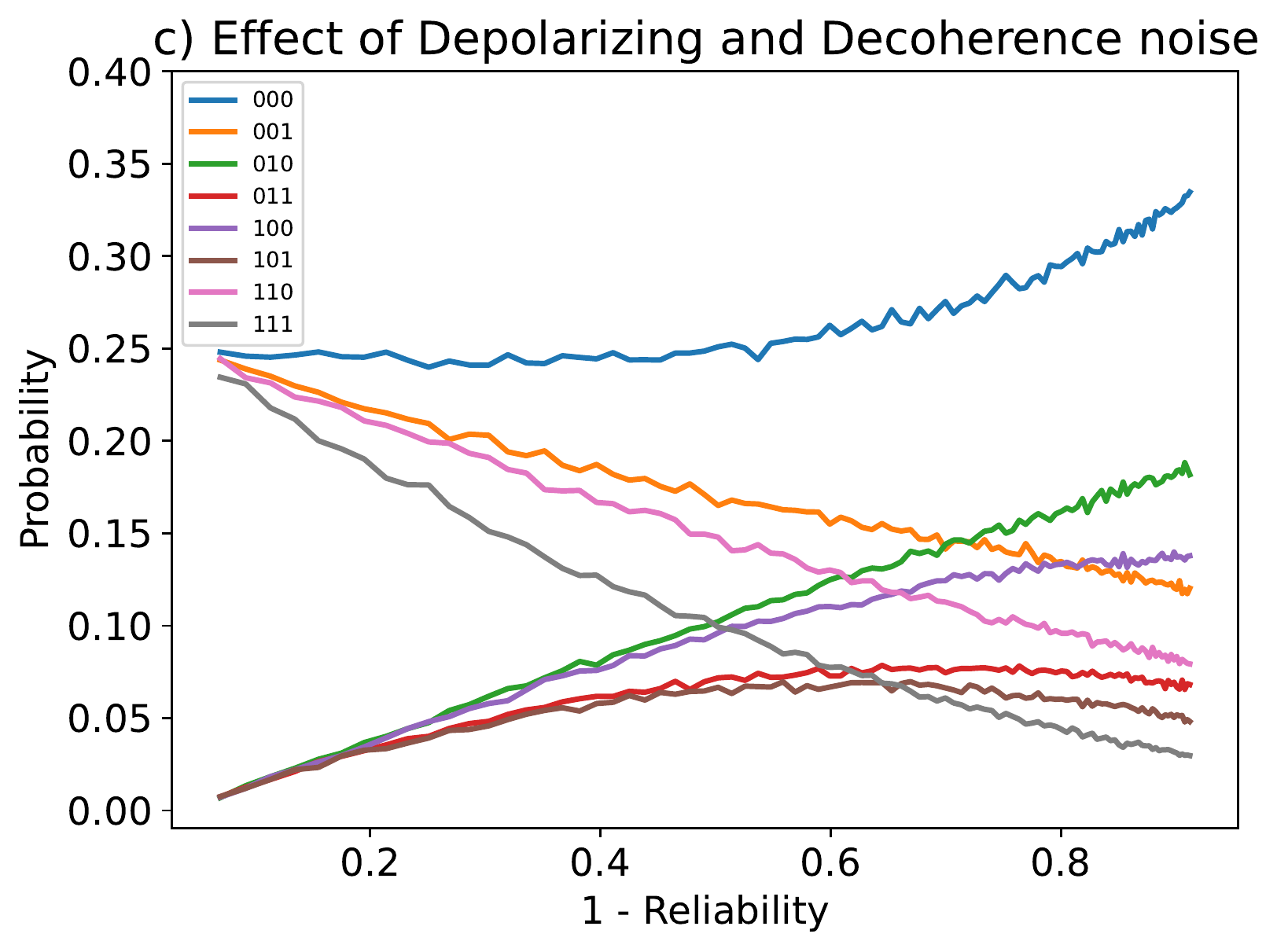}
    \caption{\centering Effect of depolarizing (a), thermal relaxation (decoherence) (b), and a combination of depolarizing and thermal relaxation errors (c) on a 3-qubit circuit. The noise free state is $\ket{\psi} = \frac{1}{2}\ket{000} + \frac{1}{2}\ket{001} + \frac{1}{2}\ket{110} + \frac{1}{2}\ket{111}$. The noise in the circuit is increased by appending pairs of CNOT gates.}
    \label{fig:1}
\end{figure*}

The depolarizing error can be modeled by the following error channel,
\begin{equation}\label{eqn:depol}
    E(\rho) = (1 - p)\rho + p\frac{I}{2^n}
\end{equation}
 where $\rho$ is the density matrix of a quantum state, $p$ is the depolarizing probability, $E(\rho)$ is noisy state after the channel, $I$ is the identity matrix, and $n$ is the number of qubits \cite{nielsen_chuang_2000}. Conceptually, a depolarizing error channel results in a weighted mixture of the noise-free quantum state and the maximally mixed state. With infinite noise, $p = 1$ and $E(\rho) = \frac{I}{2^n}$, which is the maximally mixed state. 

Thermal relaxation to the ground state is parameterized by $T_1$ and $T_2$ decoherence times \cite{9283531}. The relaxation time $T_1$ dictates the decay rate of a qubit in state $|1\rangle$ to $|0\rangle$. It is given by,
\begin{equation}\label{eqn:decoh}
    P_{\ket{1}}(t) = P_{\ket{1}}(0) e^{\frac{-t}{T_1}}
\end{equation}
where $P_{\ket{1}}(t)$ is the probability of the state $\ket{1}$ at time $t$. We can see that at $t = \infty$, % or if overall circuit latency is much longer than $T_1$, then 
the probability of the excited state approaches zero. Similarly, $T_2$ relaxation time dictates the decay rate of the phase information of a quantum state and has a similar exponential relationship.

Based on Eq. \ref{eqn:decoh}, we can further derive the decoherence impact on a multi-qubit state $\ket{q_1q_2...q_n}$. Assuming that each qubit decays independently \cite{PhysRevA.79.022108}, the decay rate would have an exponential relationship with the number of 1s (i.e., excited qubits) in the state. For example, for a 2-qubit state $\ket{11}$, the probability of it staying in $\ket{11}$ at time $t$ can be computed as follows.

\begin{equation*}
  \begin{aligned}
    P_{\ket{11}}(t) = Prob_{(q_1=\ket{1})}(t) * Prob_{(q_2=\ket{1})}(t) \\= (P_{\ket{1}}(0) e^{\frac{-t}{T_1}})(P_{\ket{1}}(0) e^{\frac{-t}{T_1}})=P_{\ket{11}}(0) e^{\frac{-2t}{T_1}}
 \end{aligned}
\end{equation*}

To illustrate the noise impact, we use a circuit to prepare a 3-qubit state $\ket{\psi} = \frac{1}{2}\ket{000} + \frac{1}{2}\ket{001} + \frac{1}{2}\ket{110} + \frac{1}{2}\ket{111}$ and then vary the noise by appending the circuit with different numbers of CNOT pairs. With a quantum noise simulator, we first only enable depolarizing noise, and the probabilities of the measured final state are shown in Fig 1a. In the figure, the $x$-axis shows the noise level, quantified with $1-reliability$, where the reliability is computed as the estimated success probability (ESP) (Eq. \ref{eqn:esp}).  
The figure shows that with more and more depolarizing noise, the 3-qubit state $\ket{\psi}$ decays to the maximally mixed state, where each 3-qubit state in ${\ket{000},\ket{001},\ket{010}, ..., \ket{111}}$ has the same probability of 1/8.

We also repeat the experiment with decoherence noise, and the results are shown in Fig. 1b. The $x$-axis is the time. We can see that with higher and higher latency resulting from the increased circuit depths, there would be stronger decoherence noise and the 3-qubit state $\ket{\psi}$ gradually decays into the $\ket{000}$ state, i.e., only $\ket{000}$ has a high probability while others have low probabilities. From Fig. \ref{fig:1}, we can see the exponential decay rate that depends on the number of 1s in a state. 

When both depolarizing and decoherence noise are enabled, the state change patterns with respect to either the noise level or latency are less obvious, as shown in Fig. 1c. Although $\ket{000}$ has a higher probability as a result of decoherence noise, other states have non-trivial probabilities due to depolarizing noise.

\subsection{Quantum Error Mitigation}
There is a plethora of QEM protocols. 
ZNE is one of the most popular QEM protocols. It works by scaling the circuit's noise and executing the circuits with different noise scales. Then it infers/extrapolates the zero-noise case from the measurement results.

Clifford Data Regression (CDR) introduced by Czarnik et al. \cite{Czarnik2021errormitigation} is a learning-based QEM. CDR uses a training set of noisy and ideal results of quantum circuits comprised entirely of Clifford gates. Simulation of such circuits is efficient on classical computers. We can hence efficiently obtain the ideal noise-free results of such quantum circuits. The set of Clifford circuits is then run on a quantum device to obtain the noisy results. Both ideal and noisy results form a training set. A linear ansatz, $a_1X^{noisy}_{|\psi\rangle}+a_2$, is deduced to correct the noisy circuit, where $a_1$ and $a_2$ are the parameters trained by machine learning methods from the training set. A recent proposal by Lowe et al.  \cite{vncdr},  variable-noise Clifford data regression (vnCDR), combines the ideas of both ZNE and CDR and achieves better mitigation results.

Introduced by Temme et al. \cite{PhysRevLett.119.180509}, Probabilistic Error Cancellation uses a linear combination of noisy quantum gates to represent ideal gates in the circuits. Quasi-probability representations are used to represent linear combinations. The representations are determined by sampling using a Monte Carlo estimation. 

A recent work, QRAFT \cite{QRAFT}, leverages the machine learning model for quantum error mitigation. With QRAFT, one executes the quantum circuit to be mitigated and then prepares a circuit appended with its inverse. In the ideal noise-free scenario, the result of the circuit and its inverse should be the initial state. However, that is not the case in a noisy scenario. The results from many such circuits (i.e., the circuit and circuit + inverse runs), along with the features of the circuits, form the training set to train a machine learning based prediction model. The prediction model then predicts the error-mitigated output based on the noisy output of a new quantum circuit.  

While the previously mentioned QEM methods are not limited by the types of error they try to mitigate, there are QEM methods that aim to mitigate a particular type of error. State Preparation and Measurement (SPAM) error mitigation methods \cite{PRXQuantum.2.040326} are such an example.  
In addition, Dynamical decoupling (DD) \cite{PhysRevLett.82.2417dd}\cite{PhysRevLett.112.050502dd} is proposed as QEM to mitigate idling errors \cite{dd_idling} as well as cross-talk errors \cite{DD_crosstalk}. 
Unlike previously mentioned QEM methods, DD does not involve running an ensemble of circuits or any post-processing step.

\section{Motivation}
The key idea of ZNE is to scale up the noise levels of a quantum circuit and then extrapolate to the zero-noise case based on the measurements of these different noise levels. One way to scale the noise is to time-scale the same unitary evolution by controlling the pulses of each gate \cite{PhysRevLett.119.180509}. Another way is to replace a unitary gate or a subcircuit or the complete circuit $U$ with ${U}({U}^\dagger{U})^n$ \cite{ giurgica2020digital}, where $n$ is the circuit scale/folding factor. 
The underlying assumption is that the noise level scales in proportion to the circuit scale/folding factor, which is accurate if the circuit noise is dominated by depolarising noise \cite{ giurgica2020digital}. Based on the assumption that errors in an operation are stochastic \cite{li2017efficient} \cite{endo2018practical}, Li and Benjamin \cite{li2017efficient} introduced Richardson extrapolation, while Endo et al. \cite{endo2018practical} introduced exponential extrapolation. Giurgica-Tiron et al. \cite{giurgica2020digital} explored various extrapolation models, such as the polynomial and the poly-exponential models, and also introduced adaptive extrapolation models.

The above-mentioned ZNE approaches have the following limitations. First, the relationship between the expectation value and the circuit scale/folding factor is nonlinear \cite{ giurgica2020digital}. As a result, different curve fitting methods have been used to extrapolate the zero-noise case, and it is difficult to decide which one should be used in practice. Second, extrapolation based on the circuit scale factor ignores the circuit complexity. For two circuits with different complexity, e.g., a single-qubit circuit with a limited depth vs. a multi-qubit circuit with a high depth, the same scaling factor may have drastically different noise impacts. %Third, ZNE requires executions of the scaled-up circuits, incurring high runtime overhead. 
Third, scaling/folding the circuit does not yield any meaningful changes in the output when the noise is high, thereby making extrapolation useless. For example, when the depolarizing noise level is very high, the output is close to the maximally mixed state. Further scaling up the noise levels would not have much impact on the output state anyway.

Next, we describe our proposed improvements to overcome these limitations.

\section{A Comprehensive Approach for Zero Noise Extrapolation}
\subsection{Overview}

Our proposed ZNE integrates different extrapolation models to mitigate different types of noise and uses quantum circuit-cut techniques \cite{CutQC,PhysRevLett.125.150504} to lower the noise level when the circuit suffers from high noise. The overall flow of our approach is shown in Fig. 2. We start by checking the circuit noise level or reliability. If the noise level/reliability is higher/lower than a threshold $\theta$, meaning that further scaling up noise levels would not lead to meaningful results, we apply circuit cut techniques to break the original circuit into subcircuits. Then, for each subcircuit, we check whether depolarizing noise is dominant. If yes, we can skip the process of decoherence noise mitigation. Otherwise, we first apply decoherence noise mitigation and then depolarizing noise mitigation. In the last step, we combine the mitigated state results of each subcircuit to produce the result for the original circuit, if the circuit was cut previously.

\begin{figure}
\centering
%\vspace*{-13mm}
{\includegraphics[width=2\linewidth]{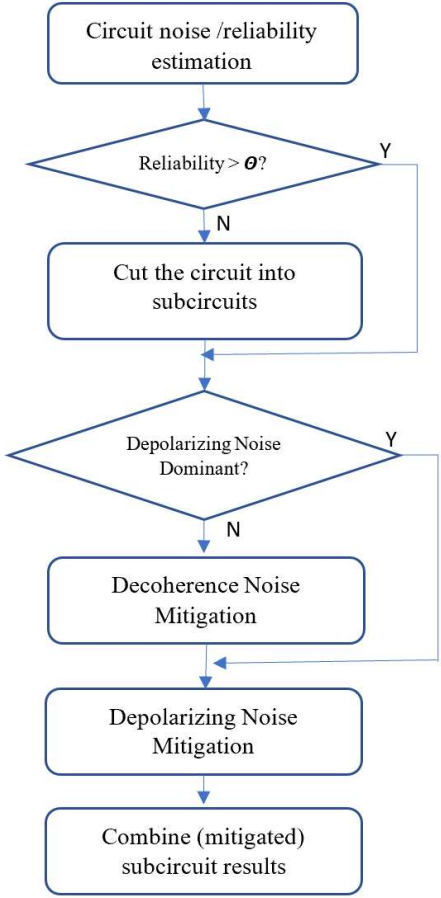}
}\hfil
\vspace*{-13mm}
\caption{Flowchart of our proposed ZNE scheme.}
\vspace*{-5mm}
\label{fig:2}
\end{figure}

Due to the different natures of depolarizing and decoherence noise, we propose different extrapolation schemes, linear extrapolation for depolarizing noise mitigation, and exponential extrapolation for decoherence noise mitigation. These two mitigation schemes are discussed in Section IV.B and IV.C, respectively. Section IV.D discusses how we scale down the noise level using the quantum circuit cut techniques.

\subsection{Reliability-based ZNE for Depolarizing Noise Mitigation}

\begin{figure}
    \centering
    \subfloat[\centering $E(\lambda)$ with different circuit folding factor $\lambda$ \label{fig:E_with_scale_factor} ]{\includegraphics[width=0.45\columnwidth]{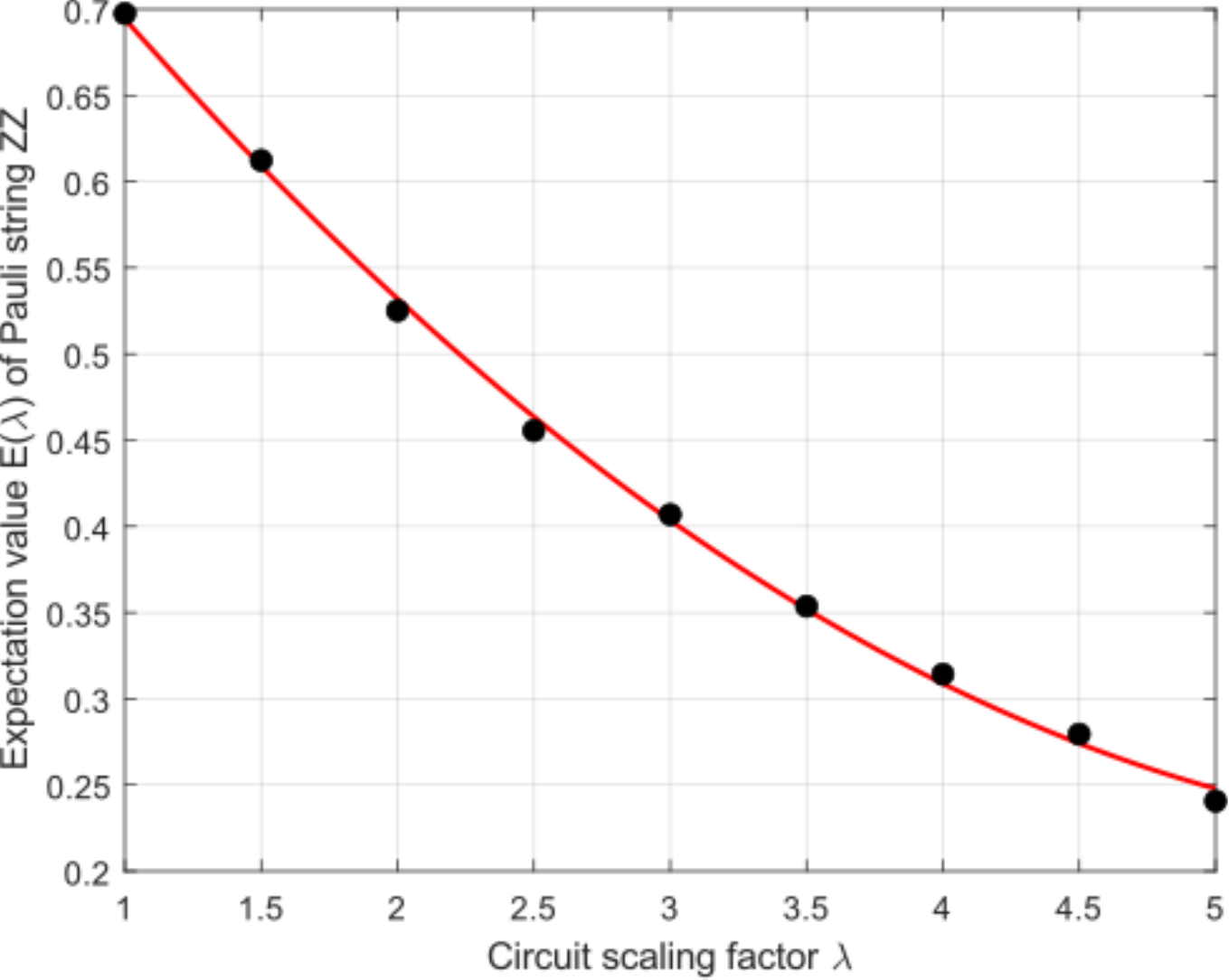}}
    \subfloat[\centering $E(\mu)$ with different reliability-based scaling factor $\mu=1-r$  \label{fig:E_with_new_scale_factor}]{\includegraphics[width=0.45\columnwidth]{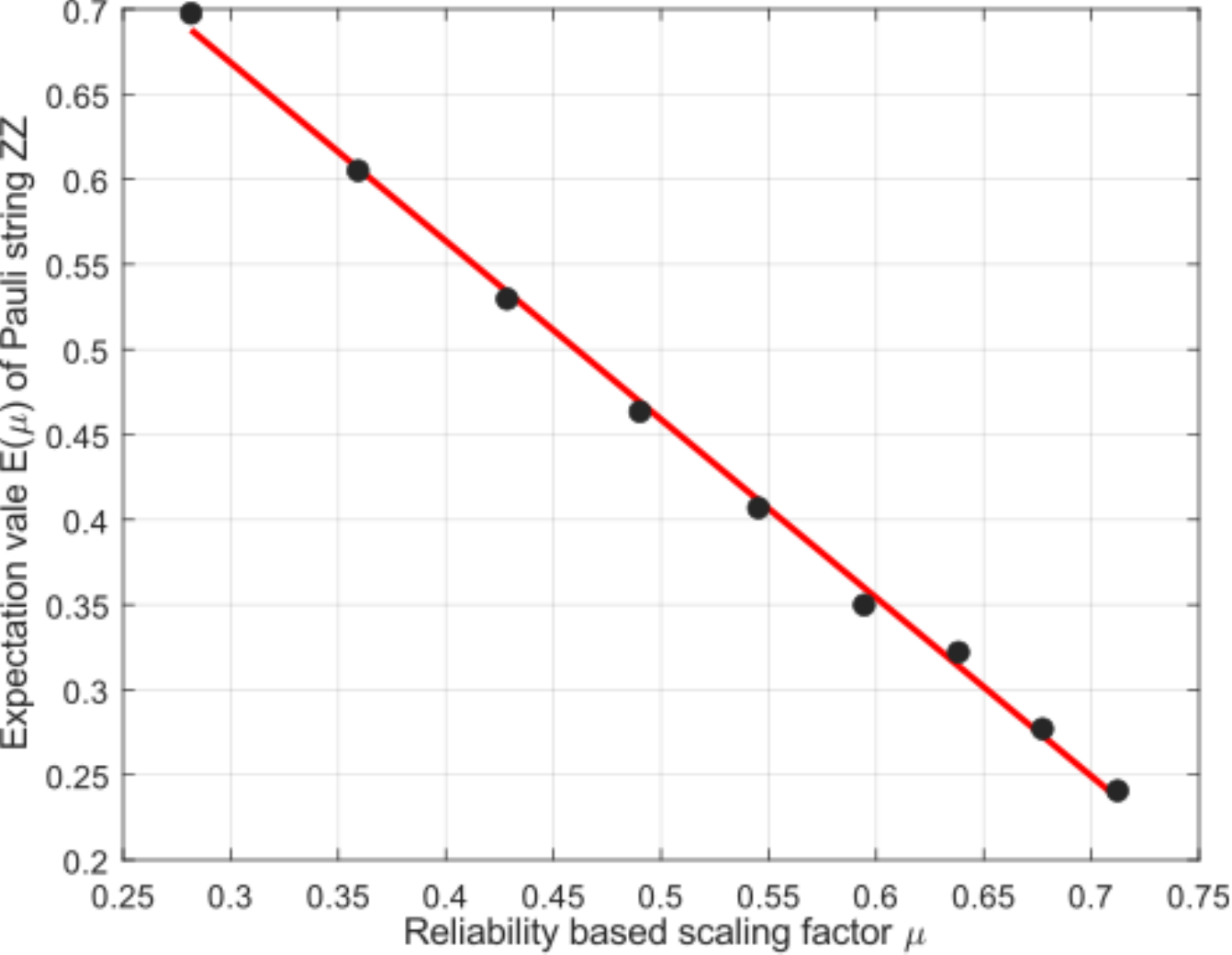}}
    \caption{Expectation values of a 2-qubit circuit with different scaling factors and the ZNE results}
    \label{fig:my_label}
\end{figure}

 We propose a novel reliability-based ZNE (RZNE) for depolarizing noise mitigation. Existing ZNE schemes such as digital ZNE \cite{giurgica2020digital} perform extrapolation based on the circuit scale factor.  In contrast, we propose to perform extrapolation based on the reliability, $r$, of the scaled quantum circuits. The motivation is to take the circuit complexity into consideration. %This way, different circuits with the same scale factor would have different reliability. 
 Since it is preferred to extrapolate to zero noise, we choose to perform extrapolation using $\mu=1-r$ rather than $r$. Zero noise means $\mu$ being 0 or $r$ being 100\%, i.e., perfect reliability. Fig.\ref{fig:E_with_scale_factor} shows an illustrating example. The y-axis is the $ZZ$ expectation values, $E(\lambda)$, of a 2-qubit quantum circuit with respect to different circuit folding factors $\lambda$. The circuits, including the original $U$ and the folded ones ${U}({U}^\dagger{U})^{(\lambda-1)}$, run on a noise simulator using the noisy model of a 5-qubit device ibmq\_quito \cite{https://quantum-computing.ibm.com/_2021}. The circuit scaling is implemented using the Mitiq framework \cite{Mitiq}. The expectation value of the noise-free execution of the circuit is 1. $E(1)$ is the expectation value based on the measurement of the original circuit and $E(0)$ is the extrapolated expectation value corresponding to the zero-noise case. In comparison, RZNE changes the $X$-axis from the circuit folding factor $\lambda$ to $\mu=1-r$, where $\mu=1-r$ can be viewed as a metric of the (reliability-based) noise scaling factor, as shown in Fig.\ref{fig:E_with_new_scale_factor}. Here, we use the simple Estimated Success Probability (ESP) \cite{nishio2020extracting} as the metric of the reliability of the quantum circuit and its noise-scaled variants. It is defined as follows. \begin{equation} \label{eqn:esp}
ESP = \prod_{i = 1}^{N_{gates}} (1-g_i^e) * \prod_{j = 0}^{N_{meas}} (1-m_j^e)
\end{equation} 
In the equation, $g_i^e$ is the gate error rate, and $m_i^e$ is the measurement error rate that are obtained from device calibration. Other metrics, such as reliability estimation using machine learning models \cite{liu2020reliability,DBLP:conf/iccad/0002LGL0J0PC022} or those considering more comprehensive noise effects \cite{arute2019quantum}, can also be used for more accurate estimation. 

A unique advantage of using reliability-based extrapolation is that it readily enables extrapolation from the extremely noisy case, which means scaling up the circuit by an infinite factor, i.e., $\lambda=\infty$. With an infinite scale factor or noise, the reliability becomes $0$ and the noise scale $\mu=1-r=1$. In comparison, such an infinite number or even a very large scale factor would be difficult to handle in curving-fitting based on $\lambda$, especially linear curves. Furthermore, the expectation value $E(\lambda=\infty)$ or $E(\mu=1)$ does not require execution of the circuit. The reason is that in this infinite noise case, the output state is the maximally mixed state when depolarizing noise is dominant, and the measured state vector would follow the uniform distribution. In our example, the $ZZ$ expectation based on the maximally mixed state or the uniformly distributed state is $\bra{\varphi}Z \otimes Z\ket{\varphi}$ = 0, if $\varphi$ is the maximally mixed state.

Another important advantage of RZNE is that a simple linear extrapolation would suffice, which we will prove next.

%\subsubsection{Linear Extrapolation from Infinite to Zero Noise}\label{sec:linear}

As shown in Eq.\ref{eqn:depol}, depolarizing error leads to a linear combination of the state itself and the maximally mixed state. The factor $p$ is the depolarizing probability. When the depolarizing noise is dominant, we can replace the term $1-p$ with the circuit reliability, $r$, and $p$ becomes $1-r$.

Hence we will have,
\begin{align}
    E(\rho) = r\rho + (1-r)\frac{I}{2^n} = r\rho + \mu\frac{I}{2^n}
\end{align}
Thus depolarizing noise is linear with respect to $r$. At infinite noise, $ r_{\infty noise} = ESP_{\infty noise} = 0$, the state is the maximally mixed state $\rho_{\infty noise} = \frac{I}{2^n}$. For the noiseless case, $ r_{zeronoise} = ESP_{zeronoise} = 1$, the state $\rho_{zeronoise} = \rho$, which can be derived as. \begin{align}
    \rho = \frac{E(\rho)}{r} - \frac{\mu}{r}\frac{I}{2^n} = \frac{E(\rho)}{r} - \frac{1-r}{r}\rho_\infty
\end{align}
where $r$ is the reliability of the original circuit, and $E(\rho)$ is the noisy result/state from the original circuit. 

\begin{figure}
\centering
\subfloat[Linear extrapolation based on I and O  \label{fig:single point rzne}]{
\begin{adjustbox}{width=0.45\linewidth}
\includegraphics[]{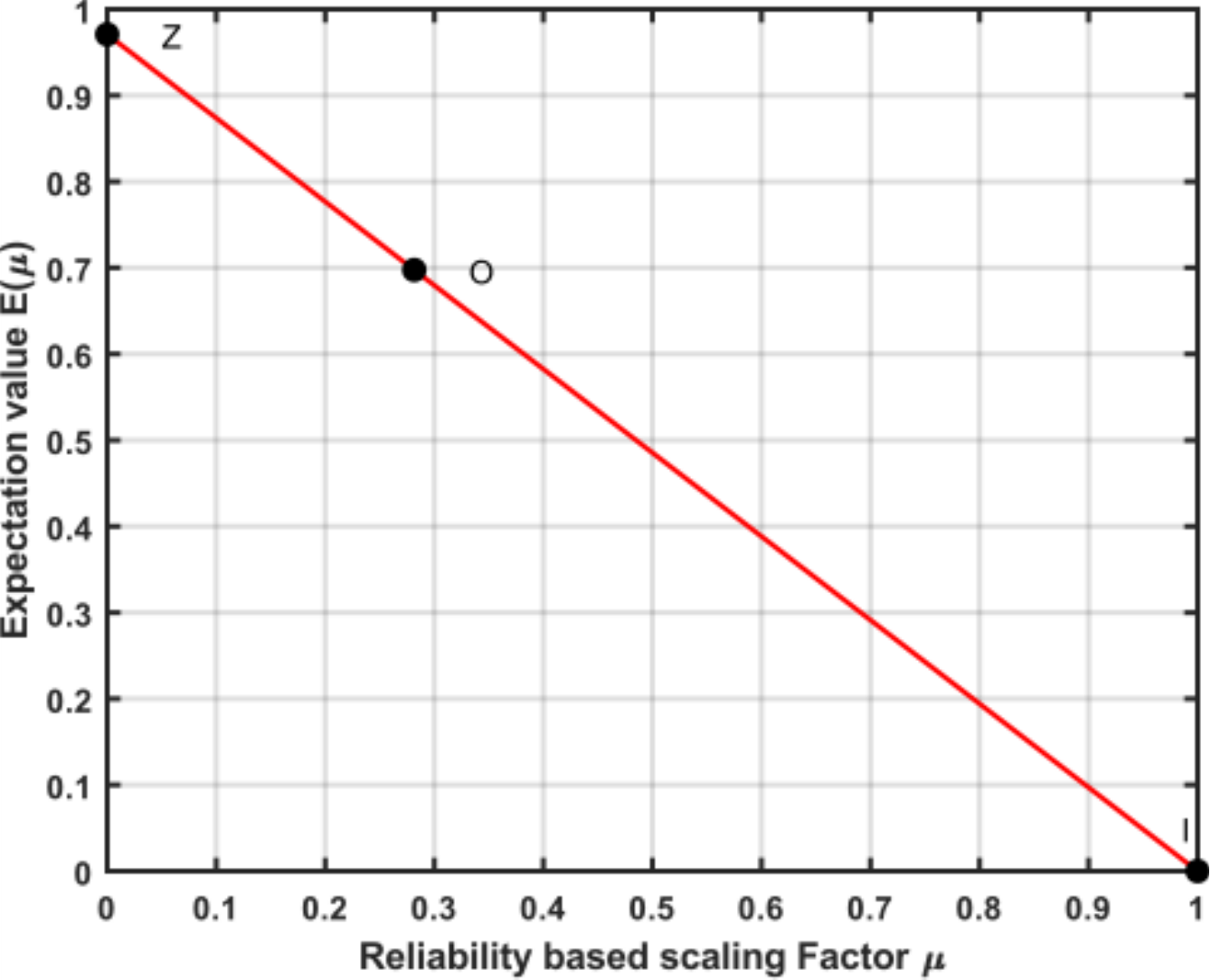}
\end{adjustbox}
}\hfil
\subfloat[Linear extrapolation based on I and multiple points of $O_j$ \label{fig:multiple points rzne}]{
\begin{adjustbox}{width=0.45\linewidth}
\begin{tikzpicture}
\begin{axis}[
    xlabel={Reliability-based scaling factor $\mu$},
    ylabel={Expectation value $E(\mu)$},
    ymajorgrids=true,
    xmajorgrids=true,
    yticklabels={,,}
    grid style=dashed,
    xmin=0, xmax=1,
    %ymin=0, %ymax=1,
]

\addplot [nodes near coords, only marks, mark = *, point meta=explicit symbolic,visualization depends on={value \thisrow{anchor}\as\myanchor},every node near coord/.append style={font=\large,anchor=\myanchor}] table [meta=label] {data3.csv};
\addplot [mark = none, color = red] {0.9025118506669925-0.9025118506669925*x};
    
\end{axis}
\end{tikzpicture}
\end{adjustbox}
}
\caption{Linear Extrapolation} 
\label{fig:two linear extrapolation}
\end{figure}
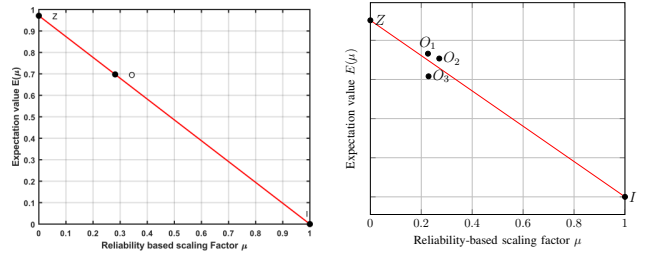

The implication of Eq. 5 is that one can extrapolate the zero noise state simply by connecting the infinite noise state and the noisy state from the original circuit, as shown in Fig.\ref{fig:single point rzne}, in which the infinite noise case is denoted as $I$, where $I_{x} = 1$ (i.e., $r=0$ or $\mu=1-r = 1$) and $I_{y} = E_{inf}$ (i.e., the expectation value based on the maximally mixed state). Similarly, the original-circuit case is denoted as $O$, and $O_{x} = r_{orig}$ (i.e., the reliability of the original circuit) and $O_{y} = E_{orig}$ (i.e., the expectation based on the measurement on the original circuit). The extrapolated zero-noise case is denoted $Z$, with $Z_{x} = 0$ (i.e., $r=1$ or $\mu=1-r = 0$) and $Z_{y} = E_{extrap}$ (i.e., the extrapolated value). RZNE can be performed either on the output state or on the expectation value. When extrapolating for the entire output state, the cost is $2^N$ extrapolations, where $N$ is the number of qubits. However, most quantum applications have few dominant states. Thus, we propose only correcting a quantum system's Top $K$ states. In other words, only the $K$ highest probability states/coefficients are extrapolated.

It is worth noting that the infinite noise case, i.e., the maximally mixed state, is independent of quantum devices or different mappings on a device. As a result, we can extend our RZNE approach to combine measurement results from different devices/mappings using linear curve fitting with the condition that the linear curve must pass the point corresponding to the infinite noise case, as illustrated in Fig. \ref{fig:multiple points rzne}, where multiple $O$ points are circuit executions on different devices or different mappings.

\RestyleAlgo{ruled}
\SetKwComment{Comment}{/* }{ */}

\begin{algorithm}[htb!]
\caption{Reliability-based ZNE (RZNE)}\label{alg:one}
%\KwData{A set $\bm{C} = \{c_1, c_2, ...c_{m}\} $, where $c_j$ is the $j^{th}$ variant of the circuit; A set $\bm{Y}$ containing the execution results corresponding to the circuits}
\KwData{The circuit $C$; ${Y_C}$ containing the execution results corresponding to the circuit}
\KwResult{An expectation value after noise mitigation}
\Begin{
  %$\bm{\mu} \gets \emptyset$\; $need\_execution = false$\;
  %\If{$\mathbf{Y} == \emptyset$}{
  %    $need\_execution = true$\;}
  %\For{$c_j \in \bm{C} $}{
    $r \gets ComputeReliability(C)$\;
    $\mu \gets 1- r$\;
    %Append($\bm{\mu},\mu $)\;
    %\tcc{Collect noisy results from circuit execution, if needed}
    %$\If{need\_execution}{
    $Y_C \gets RunCircuit(C)$\;
    %Append($\mathbf{Y},y_j $)\;}
  %}
  
  \tcc{$\mu_{m+1}$ is the factor for the infinite noise case which has reliability of $0$}
  $\mu_{m+1} \gets 1-0$\;
  \tcc{Compute expectation value based on maximally mixed state}
  $y_{m+1} \gets ComputeExpectationValue(\mu_{m+1})$\; 
  
  \tcc{$E(\mu;a,b)=a+b\mu$ is a linear model with fixed point$(\mu_{m+1}, y_{m+1})$}
  $a,b \gets BestFit(E(\mu;a,b), ({\mu},\mathbf{Y}))$\;
  \Return{$E(0;a,b)$}\;
}
\end{algorithm}
In summary, we list our RZNE algorithm in algorithm \ref{alg:one}. In the algorithm, the set $C$ is the original circuit and the same circuit with different qubit mappings or different devices, if they are used. Their reliability is computed using ESP due to its simplicity. The set $Y$ are the execution results of the circuits, i.e., either noisy states or expectation values. 
During curve fitting, the fitted linear curve is required to pass the point corresponding to the infinite noise case, $(\mu_{m+1}, y_{m+1})$. If $m=1$, the curve fitting result is simply the line connecting the infinite noise case and the noisy result of the original circuit.  The algorithm's output is the extrapolated result corresponding to zero noise or reliability being 1 (i.e., $\mu=0$). 

\subsection{State- and Latency-Aware ZNE for Decoherence Noise Mitigation}

From eq. \ref{eqn:decoh}, at $t=0$, the term $e^{\frac{-t}{T_1}}$ approaches $1$, thereby $P_{\ket{1}}(t) = P_{\ket{1}}(0)$, indicating no effect of decoherence noise. In other words, we can obtain the error-free values if we extrapolate to time $t=0$. As the decay rate depends on both the time and the number of excited 1s in a multi-qubit state, we propose the following state- and latency-aware ZNE, SLZNE, to mitigate decoherence noise. 
\begin{equation} \label{eqn:decoh_error_model}
    P_s(0) = P_s(t)/e^{\frac{-t}{T_1}}
\end{equation} 
In the equation, $t$ is the circuit latency, $c$ is the number of 'ones' in the quantum state $s$, and $P_s(t)$ is the probability of state $s$, i.e., the measured probability of the state $s$. However, the state with all zeros grows with time with probability $P_s(0) \gets  \frac{1 - P_s(t)}{e^{-t/T_1}}$.

SLZNE repeats extrapolation on every output state, which comes at the cost of $2^n$ calculations. To reduce this complexity, we resort to the same TOP $K$ approach discussed in Section IV.B. 
Our SLZNE algorithm for mitigating the decoherence noise is shown in Algorithm 2. 

\begin{algorithm}[htb!]
\caption{State- and Latency-Aware ZNE (SLZNE)}\label{alg:two}
\KwData{Circuit $C$; $T_1$ is the thermal relaxation time of the device}
\KwResult{An expectation value after noise mitigation}
\Begin{
  %$T \gets \emptyset$\;$need\_execution = false$\;
  %\If{$\mathbf{Y} == \emptyset$}{
     % $need\_execution = true$\;}

  %\For{$c_j \in \bm{C} $}{
  $t = ComputeLatency(C)$\;
  %Append($T, t_j$)\;
  %    \tcc{Collect noisy results from
%circuit execution, if needed}
    %\If{need\_execution}{
    $Y_C \gets RunCircuit(C)$\;
    %Append($\mathbf{Y},y_j $)\;%}
  %}
  
  %\tcc{$t_{m+1}$ is the latency of the infinite noise case whose latency is set as $100*T_1$}
  %$t_{m+1} \gets 100 * T_1$\;
  %\tcc{Compute expectation value based on the all zero state}
  %$y_{m+1} \gets ComputeExpectationValue(t_{m+1})$\;
  
  \tcc{Extrapolate for each state using $P_s(0)=P_s(t)/e^{-ct/T_1}$}
  \For{$s \in Y_C$}{
  \tcc{Get the number of excited qubits in state s.}
  $c \gets CalculateNumberOfOnes(s)$\;
  %\tcc{Check for all zeros states.}
  
  \If{$c == 0$}{
  $P_s(0) \gets  \frac{1 - P_s(t)}{e^{-t/T_1}}$\;
  }
  \Else{
  $P_s(0) \gets  \frac{P_s(t)}{e^{-ct/T_1}}$\;
  }
  %\EndIf
  %$P_s(0) \gets  \frac{P_s(t)}{e^{-ct/T_1}}$\;
  $Append(Y_{corrected}, P_s)$
  }
 % $z = E(0;a, b, c)$\;
    $E \gets CalculateExpectationValue(Y_{corrected})$\;
\Return{$E$}\;
}
\end{algorithm}

\subsection{Circuit-Cut based ZNE}
A quantum circuit with very low reliability has its output state close to the infinite noise case. Noise-boosting steps for such circuits result in minor changes in the circuit's output, thereby limiting the effectiveness of ZNE. 
To handle such cases, we propose to leverage circuit-cutting techniques to divide a circuit into less noisy sub-circuits. Each sub-circuit has higher reliability than the original circuit since the number of gates and/or qubits in each sub-circuit is reduced. We can apply ZNE upon each sub-circuit and then combine their results to obtain the output of the original circuit, i.e., mitgate-then-combine or CutQC-MC. 

In our scheme, we use the CutQC scheme proposed by Tang et al. \cite{CutQC} to cut the circuit into sub-circuits and then recombine the sub-circuit results. CutQC has a post-processing overhead that scales exponentially with the number of cuts. We adopt the design from \cite{CutQC}, which aims to find the cut locations to minimize this post-processing overhead. 

Our Cut-ZNE algorithm is shown in algorithm \ref{alg:three}. The input is a quantum circuit $C$. The set $S$ contains the circuit-cut results, i.e., a set of sub-circuits. If the number of cuts is lower than a threshold, $max\_num\_cuts$, we perform ZNE on each sub-circuit, first SLZNE if the decoherence noise is significant, and then RZNE for mitigating depolarizing noise. The mitigated results of each sub-circuit are combined through the $CombineResult$ function. %Both the $CutCircuit$ and $CombineResult$ functions are part of the quantum circuit cut 

\begin{algorithm}[htb!]
\caption{Cut-QC based RZNE (CutQC-MC)}\label{alg:three}
\KwData{the circuit C}
\KwResult{An error mitigated result}
\Begin{
  \tcc{$\{S\}$ is a set of subcircuits of the original circuit and cuts is the number of cuts required to separate circuit into sub circuits.}
  $\{S\}, cuts \gets CutCircuit(C)$\;
  
  \If{$cuts < max\_num\_cuts$}
  {
    $Y \gets \emptyset$\;
    \For{$s_j \in {S} $}{
    $Y_j \gets \emptyset$\;
    %$\{C_j\} \gets ScaledCircuits(\{s_j\})$\;
    $Y_{j} \gets RunCircuits(s_j)$\;
    $y_j \gets RZNE(\{s_j\}, \{y_j\})$\;
    Append(Y, $y_j$);
    }
    \Return{$ CombineResults(S, Y)$}\;
  }
} 
\end{algorithm}

Since mitigating every sub-circuit may prove costly, we also propose combine-then-mitigate (CutQC-CM), i.e., combining the noisy sub-circuit results and then mitigating the final state/expectation value. As our RZNE requires the circuit reliability for extrapolation, we use the geometric mean of all the sub-circuit reliabilities as the reliability of the combined one. The algorithm for this approach is presented in Alg. \ref{alg:four}

\begin{algorithm}[htb!]
\caption{Cut-QC based RZNE (CutQC-CM)}\label{alg:four}
\KwData{the circuit C}
\KwResult{An error mitigated result}
\Begin{
  \tcc{$\{S\}$ is a set of subcircuits of the original circuit, $\{R\}$ is their ESPs and cuts is the number of cuts required to separate circuit into sub circuits.}
  $\{S\}, cuts \gets CutCircuit(C)$\;
  $\{R\} \gets ComputeReliability(\{S\})$\;
  \If{$cuts < max\_num\_cuts$}
  {
    $Y \gets \emptyset$\;
    \For{$s_j \in {S} $}{
    $Y_j \gets \emptyset$\;
    $Y_{j} \gets RunCircuits(s_j)$\;
    Append(Y, $y_j$);
    }
    $y \gets CombineResults(S, Y)$\;
    $r \gets GeometricMean(\{R\})$\;
    \Return{$ RZNE(y, r)$}\;
  }
} 
\end{algorithm}

\section{Experimental Methodology}
\subsection{Benchmarks}
To evaluate our proposed folding-free ZNE schemes, we run the experiments on the benchmarks shown in Table I. SupermarQ \cite{9773202} is used to generate our benchmarks. 
\begin{table}[ht!]
    \centering
    \begin{tabular}{|p{0.45\columnwidth} | p{0.45\columnwidth}|}
        \hline
           Benchmark & Parameters \\
         \hline
          Hamiltonian Simulation (HS) & No. of Qubits, No. of time steps\\
          \hline
          Vanilla QAOA (QA) & No. of Qubits \\
          \hline
          Fermionic Swap QAOA (QS)  & No. of Qubits\\
          \hline
          Variational Quantum Eigensolver (VQE) & No. of Qubits, No. of Layers \\
          \hline
          GHZ State Preparation (GHZ) & No. of Qubits\\
          \hline
    \end{tabular}
    \caption{Application benchmarks.}
    \label{tab:benchmarks}
\end{table}

In our experiments, for the Top K versions of RZNE and SLZNE, K is set to a constant 10. The threshold for applying RZNE is set to $ESP > 0.10$ ($\theta =0.10$ in Fig. \ref{fig:2}), and the heuristic threshold for selectively applying SLZNE is set to $0.7 \times T_1$.

\subsection{Experimental Setup}
Our experiments are performed on the IBM Qiskit QASM simulator and a real quantum device \texttt{ibm\_cairo}. The simulator uses a noise model from IBM's 16 qubit quantum device \texttt{ibmq\_guadalupe}. The number of shots is set to 32768.

ZNE can be applied in the following two ways when the desired output is an expectation value. The first is to apply ZNE upon each individual coefficients of the output state, normalize the coefficients, and then compute the expectation value based on the noise-mitigated state. The second is to apply ZNE directly upon the expectation value, which means to extrapolate from the expectation values of the scaled circuits. In our experiments, we report the results using the first way.

We demonstrate Cut-ZNE with only the Hamiltonian Simulation, VQE and GHZ benchmark. The number of cuts required by CutQC to separate the circuits for QAOA was more than our predetermined threshold of 2. The reasoning behind our choice to limit cuts is the classical post-processing, which scales with $4^K$ with $K$ being the number of cuts. Given the exponential scaling, 2 was chosen as the maximal number of allowable cuts to demonstrate the effectiveness of Cut-ZNE. We used methods provided by authors of CutQC\cite{CutQC} for circuit cutting and sub-circuit result recombination. We show the effectiveness of approaches proposed in Alg. \ref{alg:three} and Alg. \ref{alg:four} on the aforementioned benchmarks.

We also compare our approaches with the state-of-the-art noise mitigation schemes shown in Table \ref{tab:em_methods}. For DZNE, CDR, and vnCDR, we use the software package Mitiq \cite{Mitiq}. For QRAFT, we used the code provided by the authors \cite{patel_tirthak_2021_4657000}. For DZNE, we also explore full state extrapolation, in which case the parameters are the same as in \ref{tab:em_methods}, but DZNE is applied for every element in the quantum state.
\begin{table}[ht!]
    \centering
    \begin{tabular}{|p{0.25\columnwidth} | p{0.65\columnwidth}|}
        \hline
           Method & Parameters \\
         \hline
          Digital ZNE (DZNE) & Scaling Factors = \{1, 2, 3, 4, 5\}, Extrapolation method = Linear\\
          \hline
          Clifford Data Regression (CDR) & Training set = 64 circuits, Fit method = Linear \\
          \hline
          variable noise CDR (vnCDR) & Training set = 64 circuits, Fit method = Linear, Scale Factors = {1, 2, 3, 4} \\
          \hline
          QRAFT & Training set = 1000 random circuits \\
          \hline
    \end{tabular}
    \caption{Error mitigation methods used for comparison.}
    \label{tab:em_methods}
\end{table}
\begin{comment}

\end{comment}

\subsection{Evaluation Criteria}

To evaluate the effectiveness of different QEM schemes, we use the absolute observable error (ABE) \cite{vncdr}, which is defined as follows, to quantify the error impact on the observable. %Noise-free result would have a relative error of $0$ 
\begin{equation}
    ABE(\langle O \rangle) = \abs{{\langle O \rangle_{ideal} - \langle O \rangle_{observed}}}
\end{equation}
To compute the improvement from error mitigation methods over the existing errors, we propose the following metric termed as Absolute Error Ratio (ABR),
% Need help renaming the new parameter
\begin{equation}\label{eqn:delmn}
    ABR = \abs{\frac{\langle O \rangle_{ideal} - \langle O \rangle_{mitigated}}{\langle O \rangle_{ideal} - \langle O \rangle_{noisy}}}
\end{equation}
The lower the value of ABR, the better the performance of the error mitigation method. If for a mitigation scheme, its ABR is greater than or equal to '1',  it means that the error mitigation method in question does not successfully mitigate the error compared to the unmitigated case. The reason is that the Absolute Error Ratio for an unmitigated circuit is always 1. 

For the GHZ benchmark, we report the Hellinger fidelity \cite{hellingerfidelity} instead of the ABE or ABR as GHZ does not have an observable.

\section{Experimental Results}
\begin{figure}[h]
    \centering
    \includegraphics[width = \linewidth]{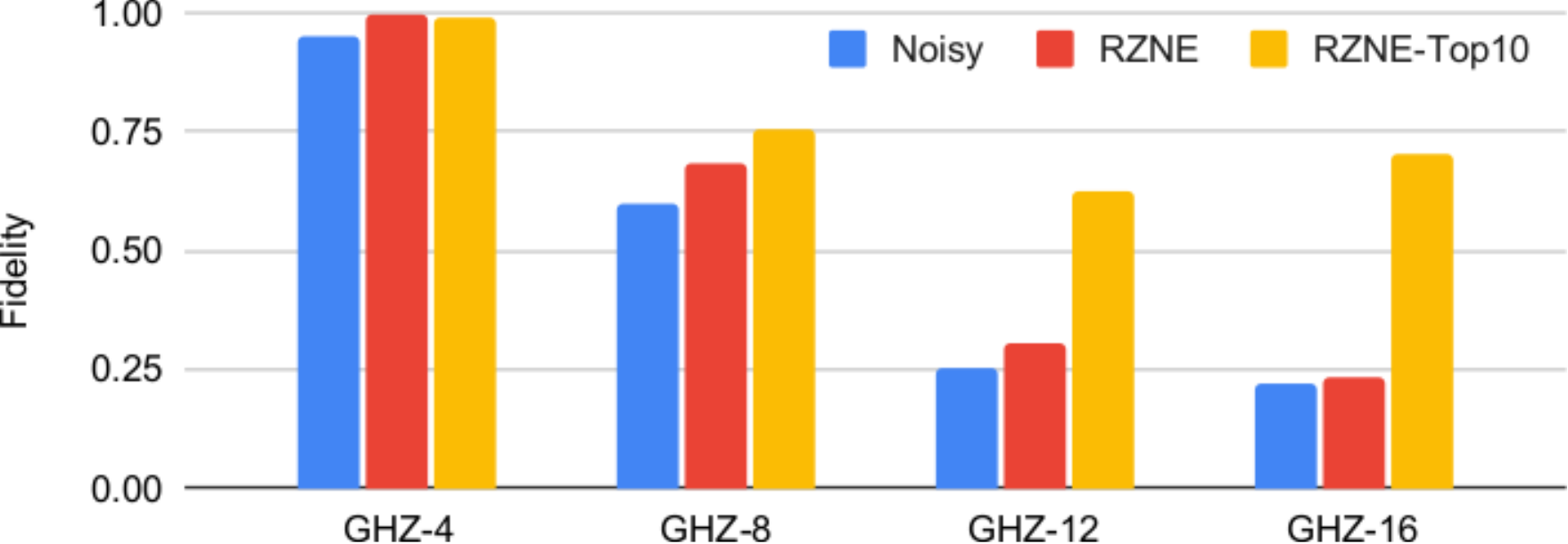}
    \caption{The fidelity of the GHZ state with depolarizing noise mitigated by RZNE and RZNE-Top 10 (Higher the better).}
    \label{fig:Depol_GHZ}
\end{figure}
\begin{figure*}[h]
    \centering
    \includegraphics[width=0.95\linewidth]{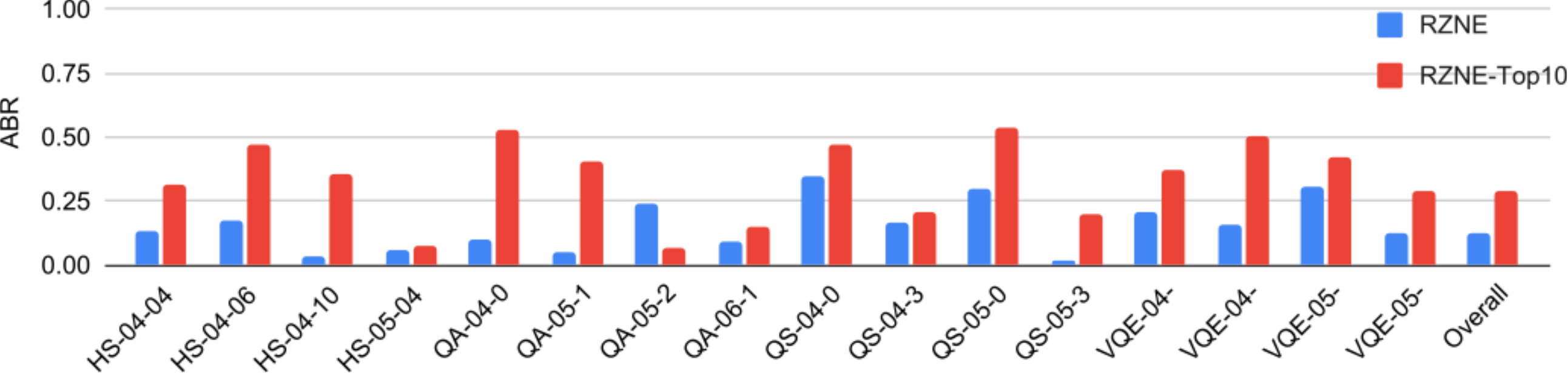}
    \caption{Absolute Error Ratios for RZNE and RZNE-Top 10 on various benchmarks under depolarizing noise. (Lower the better).}
    \label{fig:results_depol}
\end{figure*}
\subsection{Mitigating Depolarizing Noise}
We first examine the effectiveness of RZNE in mitigating errors arising from depolarizing noise. 
Depolarizing noise is simulated using Qiskit \cite{Qiskit} noise model. The noise model is constructed from the backend information of the \texttt{ibmq\_guadalupe} machine. Within the noise model, we disable the thermal relaxation error and readout error while keeping only the depolarizing error. We evaluate our method on Hamiltonian Simulation (HS), VQE, QAOA (QA), and Swap-QAOA (QS) benchmarks. Each benchmark has 4 circuits with different ESP values. For each run, we calculate the expectation value from the ideal noise-free simulation, noisy simulation, RZNE-mitigated noisy simulation, and Top 10 RZNE-mitigated simulation. As discussed in Section V.C., the absolute error ratio (ABR) is used to quantify the error impact. Our experimental results are shown in Fig. \ref{fig:results_depol} and \ref{fig:Depol_GHZ}. It is evident from the results that both variants of RZNE effectively reduce the error caused by noise for all benchmarks. On average, RZNE demonstrates an 88\% reduction in error when compared to unmitigated results, while RZNE Top 10 demonstrates a 72\% reduction. As shown in Fig. \ref{fig:results_depol}, RZNE outperforms RZNE Top 10, but requires more post-processing. The advantage of RZNE is that it corrects all the states, reducing the impact of states with higher probabilities due to noise. On the other hand, RZNE Top K only operates on a portion of the states, leaving some states uncorrected, resulting in worse performance than RZNE. %Therefore, we trade accuracy for post-processing cost, but one can easily increase the value of K for better accuracy. 
In Fig. \ref{fig:Depol_GHZ}, we observe that for the GHZ benchmarks with a higher number of qubits, RZNE Top K begins to perform significantly better than RZNE when the number of qubits increases. This is because the GHZ benchmark has two dominant states, and extrapolating only a few states prevents unintentional boosting of incorrect states. We can thus conclude that if there is prior information on the number of dominant states, the RZNE Top K method is the preferred one.

\subsection{Mitigating Decoherence Noise}
\begin{figure}[h!]
    \centering
    \includegraphics[width = \linewidth]{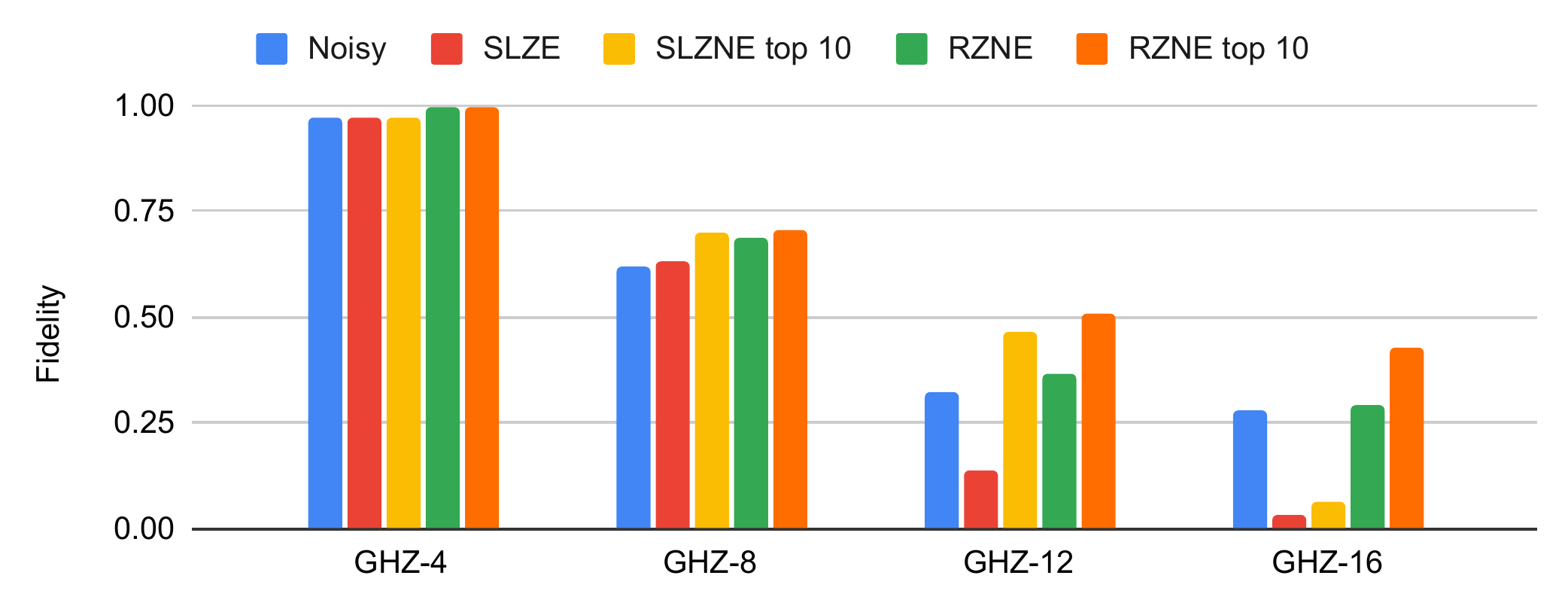}
    \caption{The fidelity of the GHZ state with decoherence noise mitigated by SLZNE and RZNE.}
    \label{fig:ghz_decoh}
\end{figure}   
\begin{figure*}[h!]
    \centering
    \includegraphics[width=\linewidth, scale=0.25]{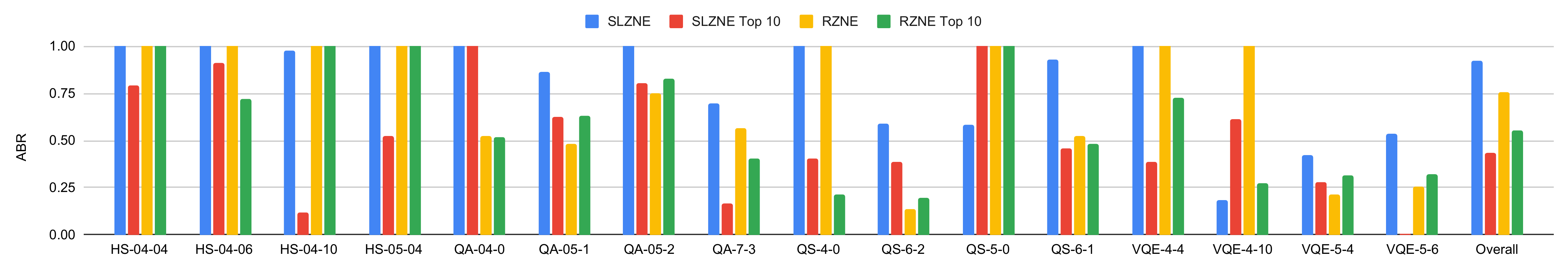}
    \caption{Absolute error ratios for SLZNE and RZNE on various benchmarks under decoherence noise.}
    \label{fig:results_decoh}
\end{figure*}

In this experiment, we examine the effectiveness of SLZNE on mitigating decoherence errors. 
We use Qiskit \cite{Qiskit} noise model with only thermal relaxation noise enabled. The median $T_1$ is $91.99 \mu s$. Fig. \ref{fig:results_decoh} and Fig. \ref{fig:ghz_decoh} report our results of SLZNE and its variant SLZNE Top-10. We also include the results using RZNE as a reference to compare the effectiveness of SLZNE.

The results in Figure \ref{fig:results_decoh} show that while all methods were successful in mitigating decoherence noise, SLZNE Top 10 is the most effective one on average. SLZNE is less optimal because it extrapolates all states, which causes some non-dominant states to be extrapolated incorrectly. When states with high number of `ones' (excited states) decay into `intermediate states', which are generally close in Hamming distance, due to decoherence. Some of these `intermediate states' may not be present in the ideal noiseless output. However, when SLZNE corrects all the states, it causes these intermediate states to incorrectly have higher probabilities. This problem is avoided by only correcting a few of the states with the highest probabilities. Then, when we re-normalize the coefficients, the intermediate states have their probabilities reduced. Another important observation is that VQE, which has higher latency, had the best performance using SLZNE, while QAOA and GHZ benchmarks with low-latency circuits resulted in poor performance of SLZNE. This can be attributed to the fact that with higher latency, the `intermediate states' are more likely to decay into the all zeros state, thereby reducing the chance that intermediate states get incorrectly mitigated. 

\subsection{Mitigating both Depolarizing and Decoherence Noise}

\begin{figure}[h!]
    \centering
    \includegraphics[width = \linewidth]{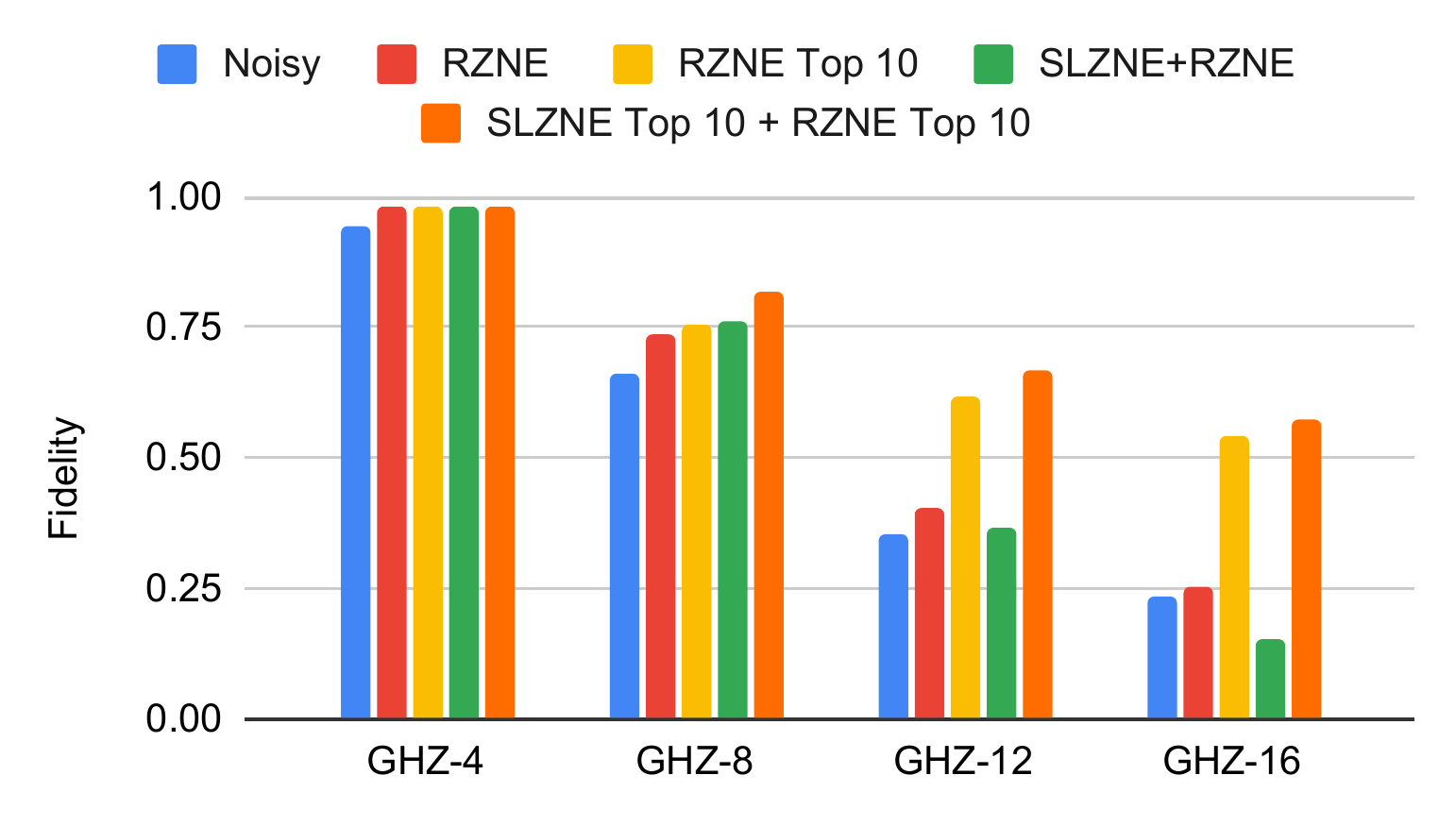}
    \caption{The fidelity of the GHZ state with deploarizing and decoherence noise mitigated by SLZNE and RZNE.}
    \label{fig:ghz_depol_decoh}
\end{figure}
\begin{figure*}[h!]
    \centering
    \includegraphics[width=\linewidth]{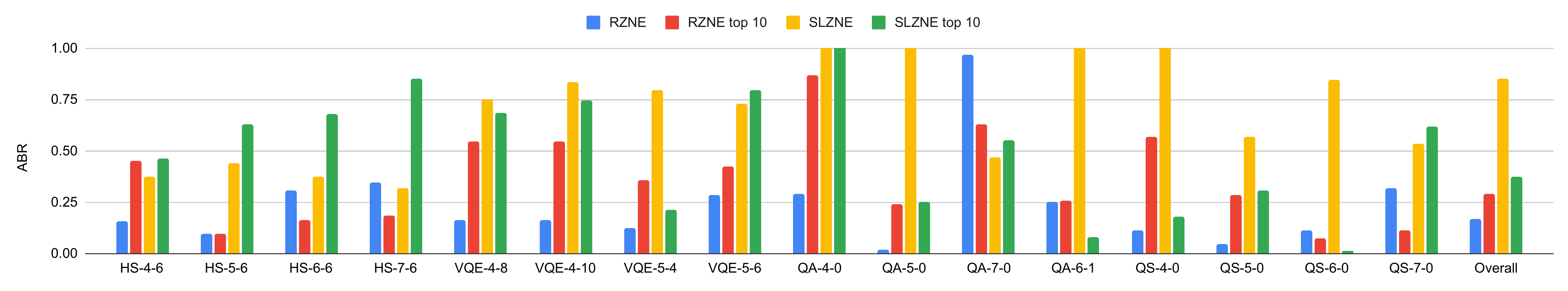}
    \caption{Absolute Error Ratios for RZNE and SLZNE on various benchmarks under decoherence and depolarizing noise.}
    \label{fig:results_decoh_depol}
    
\end{figure*}

In this experiment, both the depolarizing error and thermal relaxation error were enabled within the noise model. The results are presented in Figures \ref{fig:results_decoh_depol} and \ref{fig:ghz_depol_decoh}, which include the outcomes of both RZNE and SLZNE+RZNE, as well as their respective Top K versions. The results we obtained are consistent with previous experiments in mitigating depolarizing and decoherence noises. Specifically, we have observed that SLZNE-based methods work best for circuits with high latency such as VQE benchmark circuits, while RZNE is effective for short-depth circuits like QAOA. Based on these observations, we apply SLZNE only when the circuit latency exceeds $0.7 \times T_1$ in the remaining experiments.

\subsection{Mitigating Noise on Real Superconducting Quantum Device}
In this experiment, we perform tests on the IBM machine \texttt{ibm\_cairo}. On the real device, we reduced the number of shots to 8192. We report the observed absolute error ratios in Table \ref{tab:Realdevice}. We failed to obtain the results of other mitigation schemes since they required multiple folded circuit runs, which led to excessive wait in the job queue. Finally, having accurate estimate of gate error is important for RZNE. At times, the IBM backend may have CNOT errors equal to 1 which gives ESP to be 0. This leads to RZNE failing to extrapolate. 
\begin{table}[h]
\begin{adjustbox}{width=\columnwidth,center}
    \centering
    \begin{tabular}{|M{0.25\columnwidth}|M{0.15\columnwidth}|M{0.20\columnwidth}|M{0.20\columnwidth}|M{0.20\columnwidth}|}
    \hline
    \multirow{2}{5em}{\centering Benchmark} & \multirow{2}{5em}{\centering Qubits} & \multicolumn{1}{M{0.20\columnwidth}|}{\centering Reliability} & \multicolumn{2}{M{0.40\columnwidth}|}{\centering Absolute Error Ratio} \\ \cline{3-5}
       &  & ESP & RZNE & RZNE Top 10 \\
      \hline
      \multirow{3}{5em}{\centering Hamiltonian Simulation} & 4 & 0.654 &	0.577 &	0.027 \\
& 4 &  0.445 & 0.857 &	0.928  \\
& 7 &  0.321 & 0.566 &	0.422 \\

        \hline
        \multirow{2}{5em}{\centering VQE}  & 4 & 0.789 &	0.819 &	0.910 \\
& 4 & 0.648 & 0.292 &	0.563 \\
& 7 & 0.511 & 0.663 &	0.648 \\

      \hline
      \multirow{1}{5em}{\centering QAOA}  & 4 & 0.747 & 0.788 & 0.272 \\
      
      \hline

\end{tabular}
\end{adjustbox}
\caption{Absolute error ratios of our quantum error mitigation schemes on a real quantum device. (Lower the better)}
    \label{tab:Realdevice}
\end{table}

\begin{table}[h]
\begin{adjustbox}{width=\columnwidth,center}
    \centering
    \begin{tabular}{|M{0.16\columnwidth}|M{0.15\columnwidth}|M{0.16\columnwidth}|M{0.16\columnwidth}|M{0.16\columnwidth}|M{0.18\columnwidth}|}
    \hline
    \multirow{2}{5em}{\centering Benchmark} & \multirow{2}{5em}{\centering Qubits} & \multicolumn{1}{M{0.16\columnwidth}|}{\centering Reliability} & \multicolumn{3}{M{0.50\columnwidth}|}{\centering Fidelity} \\ \cline{3-6}
       &  & ESP & Noisy & RZNE & RZNE Top 10 \\
      \hline
      \multirow{2}{5em}{\centering GHZ} & 4 & 0.805 & 0.583 & 0.648 & 0.635   \\
      & 16 & 0.683 & 0.457 & 0.458 & 0.585 \\
      \hline
\end{tabular}
\end{adjustbox}
\caption{Fidelity of our quantum error mitigation schemes on a real quantum device. (Higher the better)}
    \label{tab:Realdevice_GHZ}
\end{table}

From Table \ref{tab:Realdevice}, we observe that our methods consistently mitigate the noises on real devices. Since \texttt{ibm\_cairo} has a relatively high $T_1$ time, we did not need to apply SLZNE.

\subsection{Comparison with the State-of-the-art QEM Approaches on Noise Simulator}

\begin{figure*}[h!]
    \centering
    \includegraphics[width=\linewidth]{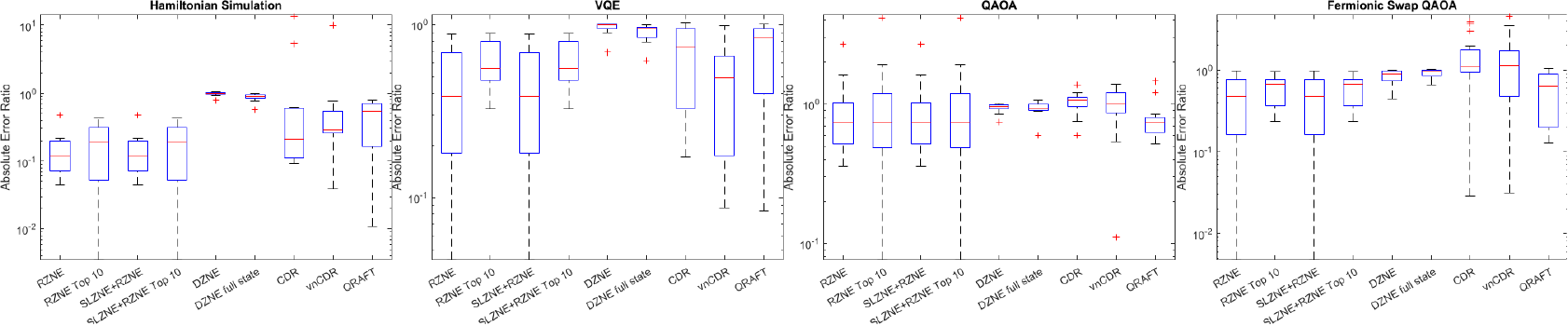}
    \caption{Comparison of RZNE and SLZNE with various State-of-Art mitigation methods. (Lower the better)}
    \label{fig:comparison}
\end{figure*}
\begin{figure}[h!]
    \centering
    \includegraphics[width =\linewidth]{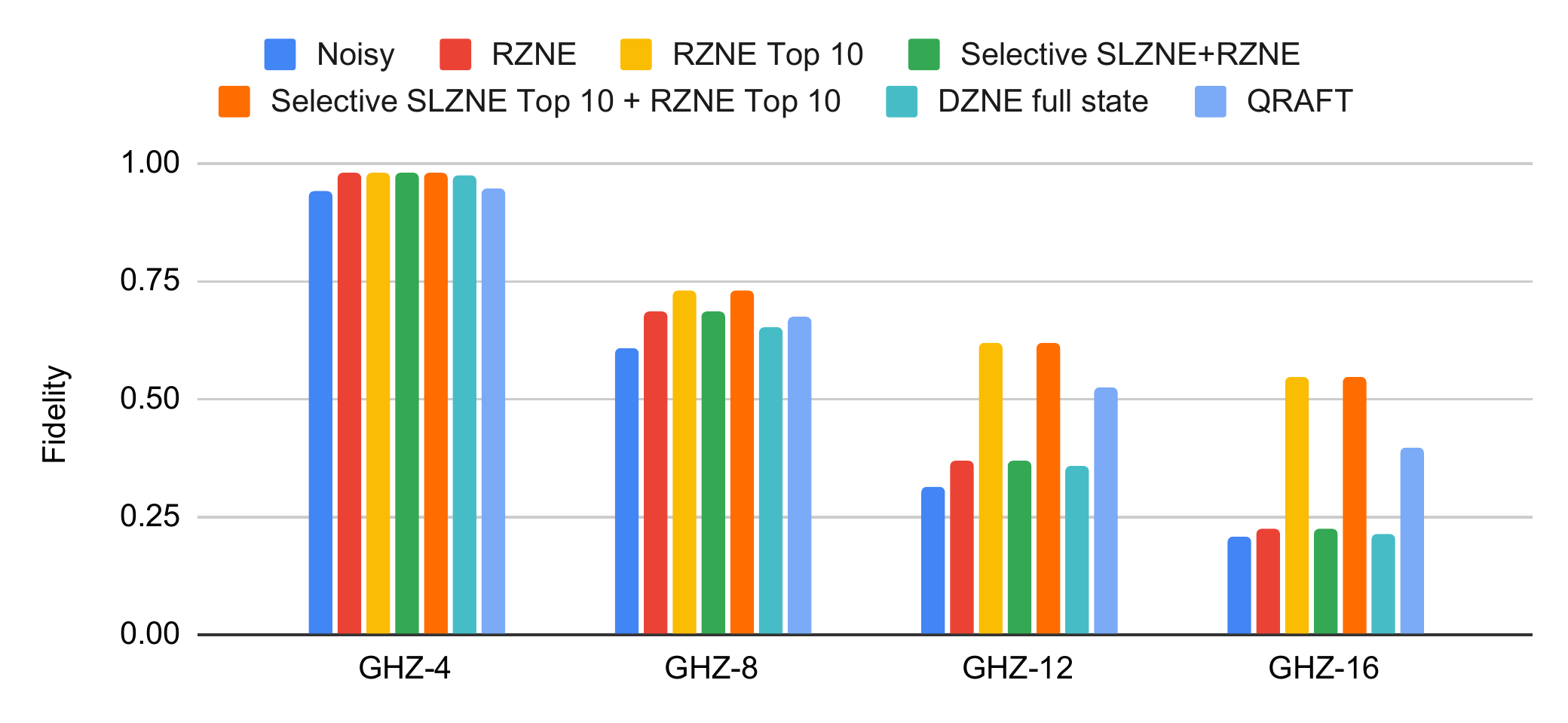}
    \caption{Comparison of error mitigation methods for the GHZ benchmark. (Higher the better).}
    \label{fig:comp_ghz}
\end{figure}

In this experiment, we test our methods against the state-of-the-art QEM methods including Digital Zero Noise Extrapolation (DZNE), Clifford Data Regression (CDR), variable noise Clifford Data Regression (vnCDR), and QRAFT. We enable all but readout error noise sources of the Qiskit noise simulator and the noise model is constructed from the backend information of the \texttt{ibmq\_guadalupe} machine. 
The experimental results are reported in Fig. \ref{fig:comparison} for the Hamiltonian Simulation, VQE, QAOA and Fermionic Swap QAOA and Fig. \ref{fig:comp_ghz} for the GHZ benchmark. Each benchmark has 16 circuits while GHZ has 4 circuits. 

We observe that RZNE consistently results in lower median error than all other state-of-the-art (SOA) methods. Our heuristic of using $0.7\times T_1$ for Selective SLZNE + RZNE can only be applied to Hamiltonian Simulation and VQE benchmarks, as only these benchmarks have circuit latencies above the heuristic threshold. In particular, RZNE performs exceptionally well for the Hamiltonian simulation benchmark compared to other error mitigation methods. The performance gain of RZNE over Digital ZNE (DZNE) is due to two reasons: 1) DZNE extrapolates over the expectation value of the observable, while RZNE extrapolates over the state; and 2) by utilizing reliability as a metric, RZNE determines how aggressively to apply the correction. It is difficult to predict the effect of noise on the expectation value of an observable. This is primarily why we need folding. We also need to choose, or find, the extrapolation model that best fits the data. However, the model that best fits the data runs a risk of overfitting leading to incorrect correction. As we cannot always guarantee prior knowledge of the correct observable, the efficacy of chosen extrapolated method cannot be guaranteed. RZNE avoids all these issues by correcting states, as we can reasonably determine the evolution of a state in the presence of specific noise factors. Thus, we can fix the choice of extrapolation method for RZNE.  Another downside of utilizing DZNE is that folding a circuit that has been significantly affected by noise does not yield any meaningful change in the value of the observable, resulting in a near-flat line for extrapolation that does not provide any scope for significant error mitigation. RZNE rectifies this by not having fixed scale factors, and for low reliability values, RZNE extrapolation is aggressive, resulting in better error mitigation.

\subsection{Results of Cut-ZNE}

\begin{table*}[]
\begin{adjustbox}{width=\textwidth,center}
    \centering
    \begin{tabular}{|c|c|c|c|c|c|c|c|c|}
    \hline
    \multirow{2}{*}{\centering Benchmark} & \multirow{2}{*}{\centering Qubits} & \multicolumn{3}{c|}{\centering Reliability} & \multicolumn{4}{c|}{\centering Absolute Error Ratio} \\ \cline{3-9}
       &  & ESP & $ESP_{subcircuit1}$ & $ESP_{subcircuit2}$  & RZNE & CutQC Unmitigated & CutQC-CM & CutQC-MC\\
      \hline
      \multirow{3}{5em}{\centering Hamiltonian Simulation} & 5 & 0.376 & 0.632 & 0.579 & 0.371 & 0.886 & 0.029 & 0.971\\
& 6 & 0.287 & 0.632 & 0.359 & 0.086 & 0.806 & 0.136 & 0.592\\
& 7 & 0.212 & 0.585 & 0.359 & 0.100 & 0.926 & 0.250 & 0.535\\
        \hline
        \multirow{3}{4em}{\centering VQE}  & 8 & 0.117 & 0.796 & 0.344 & 0.666 & 0.956 & 0.831 & 0.342 \\
& 9 & 0.087 & 0.494 & 0.344 & 0.589 & 0.663 & 0.596 & 0.399 \\
& 10 & 0.070 & 0.136 & 0.477 & 0.559 & 0.983 & 0.763 & 0.569 \\
      \hline
\end{tabular}
\end{adjustbox}
\caption{Absolute Error Ratio for Cut-ZNE error mitigation schemes. (Lower the better).}
    \label{tab:CutZNE}
\end{table*}

\begin{table*}
\begin{adjustbox}{width=\textwidth,center}
    \centering
    \begin{tabular}{|c|c|c|c|c|c|c|c|c|c|}
    \hline
      \multirow{2}{5em}{\centering Benchmark} & \multirow{2}{5em}{\centering Qubits} & \multicolumn{3}{c|}{\centering Reliability} & \multicolumn{5}{c|}{\centering Fidelity} \\ \cline{3-10}
      & & ESP & $ESP_{subcircuit1}$ & $ESP_{subcircuit2}$ & Noisy  & RZNE & CutQC Unmitigated & CutQC-CM & CutQC-MC\\
      
      \hline
      \multirow{4}{6em}{\centering GHZ} & 4 & 0.929 & 0.958 & 0.944 & 0.936 & 0.980 & 0.918 & 0.943 & 0.996 \\
& 8 & 0.604 & 0.815 & 0.930 & 0.580 & 0.662 & 0.732 & 0.764 & 0.931 \\
& 12 & 0.380 & 0.604 & 0.815 & 0.270 & 0.321 & 0.521 & 0.544 & 0.676 \\
& 16 & 0.286 & 0.737 & 0.420 & 0.213 & 0.227 & 0.363 & 0.377 & 0.432 \\
        \hline
       
\end{tabular}
\end{adjustbox}
\caption{GHZ state fidelities using Cut-ZNE error mitigation schemes. (Higher the better)}
    \label{tab:CutZNEGhz}
\end{table*}

Circuits with low ESP can be cut into smaller sub-circuits. The sub-circuits can be executed separately and the results can be recombined using classical processing to form the output state of the original circuit. This way,
their noise levels/ESPs can be reduced/increased as demonstrated in Tables \ref{tab:CutZNE} and \ref{tab:CutZNEGhz}. In these experiments, circuits were cut into 2 sub-circuits using up to 2 cuts. We apply Cut-ZNE on the Hamiltonian Simulation, VQE and GHZ benchmarks running on the Qiskit noise simulator with all noise sources but readout error enabled and the results are shown in Tables. \ref{tab:CutZNE} and \ref{tab:CutZNEGhz}. We also include the RZNE as a reference point for performance of Cut-ZNE methods. We also show the effect of simply cutting the circuit and recombining the execution results of the sub-circuits (labeled 'CutQC Unmitigated') leads to error mitigation by itself. Subsequently applying RZNE in one of two ways reduces error even further. 

In the tables, CutQC-CM (combine-then-mitigate) refers to the method in Alg. \ref{alg:four} and CutQC-MC (mitigate-then-combine) is described in Alg. \ref{alg:three}. As seen in tables all methods successfully mitigate errors as their Absolute Error Ratio is less than 1. For VQE and GHZ the best approach observed was CutQC-MC while for Hamiltonian simulation it was the CutQC-CM. RZNE proved to be better than applying CutQC for the Hamiltonian Simulation at least. Most importantly, we can see from both tables that the mitigation of errors is not just due to CutQC, and by applying our ZNE approaches, the errors are further mitigated, i.e., CutQC-CM or CutQC-MC vs. CutQC Unmitigated.

\section{Conclusions}
In this paper, we propose a comprehensive scheme for mitigating both decoherence and depolarizing noise. The key novelty includes (a) we propose reliability-based ZNE (RZNE), which performs extrapolation upon the circuit reliability such that we can easily utilize the extreme noise case and avoid folding the circuits; (b) we also show that with extrapolation upon circuit reliability, simple linear extrapolation suffices for mitigating depolarizing noise; (c) we propose SLZNE, which is an exponential extrapolation model based on circuit latency and the number of 1s in the state to mitigate decoherence noise; (d) we propose a heuristics to integrate RZNE and SLZNE: if the circuit latency is lower than $0.7\times T_1$, we use RZNE; otherwise, we first apply SLZNE then RZNE; and (e) we propose to leverage quantum circuit cut techniques to scale down the noise level so as to make ZNE approaches more effective for circuits suffering from very high noises. Our experiments on both noise simulators and real quantum devices demonstrate the effectiveness of our methods against the state-of-art quantum error mitigation schemes on commonly used benchmarks. 

\section{Acknowledgements}
 This work is partly funded by NSF grants 1818914 (with a subcontract to NC State University from Duke University), and 2120757 (with a subcontract to NC State University from the University of Maryland). It is also supported by the U.S. Department of Energy, Office of Science, National Quantum Information Science Research Centers.

%%%%%%% -- PAPER CONTENT ENDS -- %%%%%%%%

%%%%%%%%% -- BIB STYLE AND FILE -- %%%%%%%%
\bibliographystyle{IEEEtranS}
\bibliography{refs}

% Generated by IEEEtranS.bst, version: 1.13 (2008/09/30)
\begin{thebibliography}{10}
\providecommand{\url}[1]{#1}
\csname url@samestyle\endcsname
\providecommand{\newblock}{\relax}
\providecommand{\bibinfo}[2]{#2}
\providecommand{\BIBentrySTDinterwordspacing}{\spaceskip=0pt\relax}
\providecommand{\BIBentryALTinterwordstretchfactor}{4}
\providecommand{\BIBentryALTinterwordspacing}{\spaceskip=\fontdimen2\font plus
\BIBentryALTinterwordstretchfactor\fontdimen3\font minus
  \fontdimen4\font\relax}
\providecommand{\BIBforeignlanguage}[2]{{%
\expandafter\ifx\csname l@#1\endcsname\relax
\typeout{** WARNING: IEEEtranS.bst: No hyphenation pattern has been}%
\typeout{** loaded for the language `#1'. Using the pattern for}%
\typeout{** the default language instead.}%
\else
\language=\csname l@#1\endcsname
\fi
#2}}
\providecommand{\BIBdecl}{\relax}
\BIBdecl

\bibitem{hellingerfidelity}
\BIBentryALTinterwordspacing
``Hellinger fidelity.'' [Online]. Available:
  \url{https://qiskit.org/documentation/stubs/qiskit.quantum_info.hellinger_fidelity.html}
\BIBentrySTDinterwordspacing

\bibitem{patel_tirthak_2021_4657000}
\BIBentryALTinterwordspacing
\emph{QRAFT ASPLOS 21 Code and Dataset}.\hskip 1em plus 0.5em minus 0.4em\relax
  Zenodo, Feb. 2021. [Online]. Available:
  \url{https://doi.org/10.5281/zenodo.4657000}
\BIBentrySTDinterwordspacing

\bibitem{https://quantum-computing.ibm.com/_2021}
``Ibm quantum,'' https://quantum-computing.ibm.com/, 2021.

\bibitem{Qiskit}
M.~S. ANIS, Abby-Mitchell, H.~Abraham, AduOffei, R.~Agarwal, G.~Agliardi,
  M.~Aharoni, I.~Y. Akhalwaya, G.~Aleksandrowicz, T.~Alexander, M.~Amy,
  S.~Anagolum, Anthony-Gandon, E.~Arbel, A.~Asfaw, A.~Athalye, A.~Avkhadiev,
  C.~Azaustre, P.~BHOLE, A.~Banerjee, S.~Banerjee, W.~Bang, A.~Bansal,
  P.~Barkoutsos, A.~Barnawal, G.~Barron, G.~S. Barron, L.~Bello, Y.~Ben-Haim,
  M.~C. Bennett, D.~Bevenius, D.~Bhatnagar, A.~Bhobe, P.~Bianchini, L.~S.
  Bishop, C.~Blank, S.~Bolos, S.~Bopardikar, S.~Bosch, S.~Brandhofer, Brandon,
  S.~Bravyi, N.~Bronn, Bryce-Fuller, D.~Bucher, A.~Burov, F.~Cabrera,
  P.~Calpin, L.~Capelluto, J.~Carballo, G.~Carrascal, A.~Carriker, I.~Carvalho,
  A.~Chen, C.-F. Chen, E.~Chen, J.~C. Chen, R.~Chen, F.~Chevallier, K.~Chinda,
  R.~Cholarajan, J.~M. Chow, S.~Churchill, CisterMoke, C.~Claus, C.~Clauss,
  C.~Clothier, R.~Cocking, R.~Cocuzzo, J.~Connor, F.~Correa, Z.~Crockett, A.~J.
  Cross, A.~W. Cross, S.~Cross, J.~Cruz-Benito, C.~Culver, A.~D.
  C{\'o}rcoles-Gonzales, N.~D, S.~Dague, T.~E. Dandachi, A.~N. Dangwal,
  J.~Daniel, M.~Daniels, M.~Dartiailh, A.~R. Davila, F.~Debouni, A.~Dekusar,
  A.~Deshmukh, M.~Deshpande, D.~Ding, J.~Doi, E.~M. Dow, P.~Downing,
  E.~Drechsler, E.~Dumitrescu, K.~Dumon, I.~Duran, K.~EL-Safty, E.~Eastman,
  G.~Eberle, A.~Ebrahimi, P.~Eendebak, D.~Egger, ElePT, Emilio, A.~Espiricueta,
  M.~Everitt, D.~Facoetti, Farida, P.~M. Fern{\'a}ndez, S.~Ferracin,
  D.~Ferrari, A.~H. Ferrera, R.~Fouilland, A.~Frisch, A.~Fuhrer, B.~Fuller,
  M.~GEORGE, J.~Gacon, B.~G. Gago, C.~Gambella, J.~M. Gambetta, A.~Gammanpila,
  L.~Garcia, T.~Garg, S.~Garion, J.~R. Garrison, J.~Garrison, T.~Gates,
  H.~Georgiev, L.~Gil, A.~Gilliam, A.~Giridharan, J.~Gomez-Mosquera, Gonzalo,
  S.~de~la Puente~Gonz{\'a}lez, J.~Gorzinski, I.~Gould, D.~Greenberg,
  D.~Grinko, W.~Guan, D.~Guijo, J.~A. Gunnels, H.~Gupta, N.~Gupta, J.~M.
  G{\"u}nther, M.~Haglund, I.~Haide, I.~Hamamura, O.~C. Hamido, F.~Harkins,
  K.~Hartman, A.~Hasan, V.~Havlicek, J.~Hellmers, {\L}.~Herok, S.~Hillmich,
  H.~Horii, C.~Howington, S.~Hu, W.~Hu, J.~Huang, R.~Huisman, H.~Imai,
  T.~Imamichi, K.~Ishizaki, Ishwor, R.~Iten, T.~Itoko, A.~Ivrii, A.~Javadi,
  A.~Javadi-Abhari, W.~Javed, Q.~Jianhua, M.~Jivrajani, K.~Johns, S.~Johnstun,
  Jonathan-Shoemaker, JosDenmark, JoshDumo, J.~Judge, T.~Kachmann, A.~Kale,
  N.~Kanazawa, J.~Kane, Kang-Bae, A.~Kapila, A.~Karazeev, P.~Kassebaum,
  T.~Kehrer, J.~Kelso, S.~Kelso, V.~Khanderao, S.~King, Y.~Kobayashi,
  Kovi11Day, A.~Kovyrshin, R.~Krishnakumar, V.~Krishnan, K.~Krsulich,
  P.~Kumkar, G.~Kus, R.~LaRose, E.~Lacal, R.~Lambert, H.~Landa, J.~Lapeyre,
  J.~Latone, S.~Lawrence, C.~Lee, G.~Li, J.~Lishman, D.~Liu, P.~Liu, Lolcroc,
  A.~K. M, L.~Madden, Y.~Maeng, S.~Maheshkar, K.~Majmudar, A.~Malyshev, M.~E.
  Mandouh, J.~Manela, Manjula, J.~Marecek, M.~Marques, K.~Marwaha, D.~Maslov,
  P.~Maszota, D.~Mathews, A.~Matsuo, F.~Mazhandu, D.~McClure, M.~McElaney,
  C.~McGarry, D.~McKay, D.~McPherson, S.~Meesala, D.~Meirom, C.~Mendell,
  T.~Metcalfe, M.~Mevissen, A.~Meyer, A.~Mezzacapo, R.~Midha, D.~Miller,
  Z.~Minev, A.~Mitchell, N.~Moll, A.~Montanez, G.~Monteiro, M.~D. Mooring,
  R.~Morales, N.~Moran, D.~Morcuende, S.~Mostafa, M.~Motta, R.~Moyard,
  P.~Murali, D.~Murata, J.~M{\"u}ggenburg, T.~NEMOZ, D.~Nadlinger,
  K.~Nakanishi, G.~Nannicini, P.~Nation, E.~Navarro, Y.~Naveh, S.~W. Neagle,
  P.~Neuweiler, A.~Ngoueya, T.~Nguyen, J.~Nicander, Nick-Singstock, P.~Niroula,
  H.~Norlen, NuoWenLei, L.~J. O'Riordan, O.~Ogunbayo, P.~Ollitrault,
  T.~Onodera, R.~Otaolea, S.~Oud, D.~Padilha, H.~Paik, S.~Pal, Y.~Pang,
  A.~Panigrahi, V.~R. Pascuzzi, S.~Perriello, E.~Peterson, A.~Phan, K.~Pilch,
  F.~Piro, M.~Pistoia, C.~Piveteau, J.~Plewa, P.~Pocreau, A.~Pozas-Kerstjens,
  R.~Pracht, M.~Prokop, V.~Prutyanov, S.~Puri, D.~Puzzuoli, J.~P{\'e}rez,
  Quant02, Quintiii, R.~I. Rahman, A.~Raja, R.~Rajeev, I.~Rajput, N.~Ramagiri,
  A.~Rao, R.~Raymond, O.~Reardon-Smith, R.~M.-C. Redondo, M.~Reuter, J.~Rice,
  M.~Riedemann, Rietesh, D.~Risinger, M.~L. Rocca, D.~M. Rodr{\'\i}guez,
  RohithKarur, B.~Rosand, M.~Rossmannek, M.~Ryu, T.~SAPV, N.~R.~C. Sa, A.~Saha,
  A.~Ash-Saki, S.~Sanand, M.~Sandberg, H.~Sandesara, R.~Sapra, H.~Sargsyan,
  A.~Sarkar, N.~Sathaye, B.~Schmitt, C.~Schnabel, Z.~Schoenfeld, T.~L.
  Scholten, E.~Schoute, M.~Schulterbrandt, J.~Schwarm, J.~Seaward, Sergi, I.~F.
  Sertage, K.~Setia, F.~Shah, N.~Shammah, R.~Sharma, Y.~Shi, J.~Shoemaker,
  A.~Silva, A.~Simonetto, D.~Singh, D.~Singh, P.~Singh, P.~Singkanipa,
  Y.~Siraichi, Siri, J.~Sistos, I.~Sitdikov, S.~Sivarajah, Slavikmew, M.~B.
  Sletfjerding, J.~A. Smolin, M.~Soeken, I.~O. Sokolov, I.~Sokolov, V.~P.
  Soloviev, SooluThomas, Starfish, D.~Steenken, M.~Stypulkoski, A.~Suau,
  S.~Sun, K.~J. Sung, M.~Suwama, O.~S{\l}owik, H.~Takahashi, T.~Takawale,
  I.~Tavernelli, C.~Taylor, P.~Taylour, S.~Thomas, K.~Tian, M.~Tillet, M.~Tod,
  M.~Tomasik, C.~Tornow, E.~de~la Torre, J.~L.~S. Toural, K.~Trabing,
  M.~Treinish, D.~Trenev, TrishaPe, F.~Truger, G.~Tsilimigkounakis, D.~Tulsi,
  W.~Turner, Y.~Vaknin, C.~R. Valcarce, F.~Varchon, A.~Vartak, A.~C. Vazquez,
  P.~Vijaywargiya, V.~Villar, B.~Vishnu, D.~Vogt-Lee, C.~Vuillot, J.~Weaver,
  J.~Weidenfeller, R.~Wieczorek, J.~A. Wildstrom, J.~Wilson, E.~Winston,
  WinterSoldier, J.~J. Woehr, S.~Woerner, R.~Woo, C.~J. Wood, R.~Wood, S.~Wood,
  J.~Wootton, M.~Wright, L.~Xing, J.~YU, B.~Yang, U.~Yang, J.~Yao, D.~Yeralin,
  R.~Yonekura, D.~Yonge-Mallo, R.~Yoshida, R.~Young, J.~Yu, L.~Yu, C.~Zachow,
  L.~Zdanski, H.~Zhang, I.~Zidaru, B.~Zimmermann, C.~Zoufal, aeddins ibm,
  alexzhang13, b63, bartek bartlomiej, bcamorrison, brandhsn, charmerDark,
  deeplokhande, dekel.meirom, dime10, dlasecki, ehchen, fanizzamarco,
  fs1132429, gadial, galeinston, georgezhou20, georgios ts, gruu, hhorii,
  hykavitha, itoko, jeppevinkel, jessica angel7, jezerjojo14, jliu45, jscott2,
  klinvill, krutik2966, ma5x, michelle4654, msuwama, nico lgrs, nrhawkins,
  ntgiwsvp, ordmoj, sagar pahwa, pritamsinha2304, ryancocuzzo, saktar unr,
  saswati qiskit, septembrr, sethmerkel, sg495, shaashwat, smturro2,
  sternparky, strickroman, tigerjack, tsura crisaldo, upsideon, vadebayo49,
  welien, willhbang, wmurphy collabstar, yang.luh, and M.~{\v{C}}epulkovskis,
  ``Qiskit: An open-source framework for quantum computing,'' 2021.

\bibitem{arute2019quantum}
F.~Arute, K.~Arya, R.~Babbush, D.~Bacon, J.~C. Bardin, R.~Barends, R.~Biswas,
  S.~Boixo, F.~G. Brandao, D.~A. Buell \emph{et~al.}, ``Quantum supremacy using
  a programmable superconducting processor,'' \emph{Nature}, vol. 574, no.
  7779, pp. 505--510, 2019.

\bibitem{PhysRevA.79.022108}
\BIBentryALTinterwordspacing
A.~Borras, A.~P. Majtey, A.~R. Plastino, M.~Casas, and A.~Plastino,
  ``Robustness of highly entangled multiqubit states under decoherence,''
  \emph{Phys. Rev. A}, vol.~79, p. 022108, Feb 2009. [Online]. Available:
  \url{https://link.aps.org/doi/10.1103/PhysRevA.79.022108}
\BIBentrySTDinterwordspacing

\bibitem{Czarnik2021errormitigation}
\BIBentryALTinterwordspacing
P.~Czarnik, A.~Arrasmith, P.~J. Coles, and L.~Cincio, ``Error mitigation with
  {C}lifford quantum-circuit data,'' \emph{{Quantum}}, vol.~5, p. 592, Nov.
  2021. [Online]. Available: \url{https://doi.org/10.22331/q-2021-11-26-592}
\BIBentrySTDinterwordspacing

\bibitem{dd_idling}
\BIBentryALTinterwordspacing
P.~Das, S.~Tannu, S.~Dangwal, and M.~Qureshi, ``Adapt: Mitigating idling errors
  in qubits via adaptive dynamical decoupling,'' in \emph{MICRO-54: 54th Annual
  IEEE/ACM International Symposium on Microarchitecture}, ser. MICRO '21.\hskip
  1em plus 0.5em minus 0.4em\relax New York, NY, USA: Association for Computing
  Machinery, 2021, p. 950–962. [Online]. Available:
  \url{https://doi.org/10.1145/3466752.3480059}
\BIBentrySTDinterwordspacing

\bibitem{endo2018practical}
S.~Endo, S.~C. Benjamin, and Y.~Li, ``Practical quantum error mitigation for
  near-future applications,'' \emph{Physical Review X}, vol.~8, no.~3, p.
  031027, 2018.

\bibitem{giurgica2020digital}
T.~Giurgica-Tiron, Y.~Hindy, R.~LaRose, A.~Mari, and W.~J. Zeng, ``Digital zero
  noise extrapolation for quantum error mitigation,'' in \emph{QCE}, 2020.

\bibitem{Mitiq}
R.~LaRose, A.~Mari, S.~Kaiser, P.~J. Karalekas, A.~A. Alves, P.~Czarnik, M.~E.
  Mandouh, M.~H. Gordon, Y.~Hindy, A.~Robertson, P.~Thakre, N.~Shammah, and
  W.~J. Zeng, ``Mitiq: A software package for error mitigation on noisy quantum
  computers,'' 2020.

\bibitem{li2017efficient}
Y.~Li and S.~C. Benjamin, ``Efficient variational quantum simulator
  incorporating active error minimization,'' \emph{Physical Review X}, vol.~7,
  no.~2, 2017.

\bibitem{liu2020reliability}
J.~Liu and H.~Zhou, ``Reliability modeling of nisq-era quantum computers,'' in
  \emph{IISWC}, 2020.

\bibitem{vncdr}
\BIBentryALTinterwordspacing
A.~Lowe, M.~H. Gordon, P.~Czarnik, A.~Arrasmith, P.~J. Coles, and L.~Cincio,
  ``Unified approach to data-driven quantum error mitigation,'' \emph{Phys.
  Rev. Research}, vol.~3, p. 033098, Jul 2021. [Online]. Available:
  \url{https://link.aps.org/doi/10.1103/PhysRevResearch.3.033098}
\BIBentrySTDinterwordspacing

\bibitem{crosstalk_scheduling}
\BIBentryALTinterwordspacing
P.~Murali, D.~C. Mckay, M.~Martonosi, and A.~Javadi-Abhari, ``Software
  mitigation of crosstalk on noisy intermediate-scale quantum computers,'' ser.
  ASPLOS '20.\hskip 1em plus 0.5em minus 0.4em\relax New York, NY, USA:
  Association for Computing Machinery, 2020, p. 1001–1016. [Online].
  Available: \url{https://doi.org/10.1145/3373376.3378477}
\BIBentrySTDinterwordspacing

\bibitem{PRXQuantum.2.040326}
\BIBentryALTinterwordspacing
P.~D. Nation, H.~Kang, N.~Sundaresan, and J.~M. Gambetta, ``Scalable mitigation
  of measurement errors on quantum computers,'' \emph{PRX Quantum}, vol.~2, p.
  040326, Nov 2021. [Online]. Available:
  \url{https://link.aps.org/doi/10.1103/PRXQuantum.2.040326}
\BIBentrySTDinterwordspacing

\bibitem{nielsen_chuang_2000}
M.~A. Nielsen and I.~L. Chuang, \emph{Quantum Computation and Quantum
  Information}.\hskip 1em plus 0.5em minus 0.4em\relax Cambridge University
  Press, 2000.

\bibitem{nishio2020extracting}
S.~Nishio, Y.~Pan, T.~Satoh, H.~Amano, and R.~V. Meter, ``Extracting success
  from ibm’s 20-qubit machines using error-aware compilation,'' \emph{ACM
  Journal on Emerging Technologies in Computing Systems (JETC)}, vol.~16,
  no.~3, pp. 1--25, 2020.

\bibitem{QRAFT}
\BIBentryALTinterwordspacing
T.~Patel and D.~Tiwari, ``Qraft: Reverse your quantum circuit and know the
  correct program output,'' in \emph{Proceedings of the 26th ACM International
  Conference on Architectural Support for Programming Languages and Operating
  Systems}, ser. ASPLOS '21.\hskip 1em plus 0.5em minus 0.4em\relax New York,
  NY, USA: Association for Computing Machinery, 2021, p. 443–455. [Online].
  Available: \url{https://doi.org/10.1145/3445814.3446743}
\BIBentrySTDinterwordspacing

\bibitem{PhysRevLett.125.150504}
\BIBentryALTinterwordspacing
T.~Peng, A.~W. Harrow, M.~Ozols, and X.~Wu, ``Simulating large quantum circuits
  on a small quantum computer,'' \emph{Phys. Rev. Lett.}, vol. 125, p. 150504,
  Oct 2020. [Online]. Available:
  \url{https://link.aps.org/doi/10.1103/PhysRevLett.125.150504}
\BIBentrySTDinterwordspacing

\bibitem{CutQC}
\BIBentryALTinterwordspacing
W.~Tang, T.~Tomesh, M.~Suchara, J.~Larson, and M.~Martonosi, ``Cutqc: Using
  small quantum computers for large quantum circuit evaluations,'' in
  \emph{Proceedings of the 26th ACM International Conference on Architectural
  Support for Programming Languages and Operating Systems}, ser. ASPLOS
  '21.\hskip 1em plus 0.5em minus 0.4em\relax New York, NY, USA: Association
  for Computing Machinery, 2021, p. 473–486. [Online]. Available:
  \url{https://doi.org/10.1145/3445814.3446758}
\BIBentrySTDinterwordspacing

\bibitem{PhysRevLett.119.180509}
\BIBentryALTinterwordspacing
K.~Temme, S.~Bravyi, and J.~M. Gambetta, ``Error mitigation for short-depth
  quantum circuits,'' \emph{Phys. Rev. Lett.}, vol. 119, Nov 2017. [Online].
  Available: \url{https://link.aps.org/doi/10.1103/PhysRevLett.119.180509}
\BIBentrySTDinterwordspacing

\bibitem{9773202}
\BIBentryALTinterwordspacing
T.~Tomesh, P.~Gokhale, V.~Omole, G.~Ravi, K.~N. Smith, J.~Viszlai, X.~Wu,
  N.~Hardavellas, M.~R. Martonosi, and F.~T. Chong, ``Supermarq: A scalable
  quantum benchmark suite,'' in \emph{2022 IEEE International Symposium on
  High-Performance Computer Architecture (HPCA)}.\hskip 1em plus 0.5em minus
  0.4em\relax Los Alamitos, CA, USA: IEEE Computer Society, apr 2022, pp.
  587--603. [Online]. Available:
  \url{https://doi.ieeecomputersociety.org/10.1109/HPCA53966.2022.00050}
\BIBentrySTDinterwordspacing

\bibitem{DD_crosstalk}
\BIBentryALTinterwordspacing
V.~Tripathi, H.~Chen, M.~Khezri, K.-W. Yip, E.~Levenson-Falk, and D.~A. Lidar,
  ``Suppression of crosstalk in superconducting qubits using dynamical
  decoupling,'' \emph{Phys. Rev. Applied}, vol.~18, p. 024068, Aug 2022.
  [Online]. Available:
  \url{https://link.aps.org/doi/10.1103/PhysRevApplied.18.024068}
\BIBentrySTDinterwordspacing

\bibitem{PhysRevLett.82.2417dd}
\BIBentryALTinterwordspacing
L.~Viola, E.~Knill, and S.~Lloyd, ``Dynamical decoupling of open quantum
  systems,'' \emph{Phys. Rev. Lett.}, vol.~82, pp. 2417--2421, Mar 1999.
  [Online]. Available:
  \url{https://link.aps.org/doi/10.1103/PhysRevLett.82.2417}
\BIBentrySTDinterwordspacing

\bibitem{DBLP:conf/iccad/0002LGL0J0PC022}
\BIBentryALTinterwordspacing
H.~Wang, Z.~Liang, J.~Gu, Z.~Li, Y.~Ding, W.~Jiang, Y.~Shi, D.~Z. Pan, F.~T.
  Chong, and S.~Han, ``Torchquantum case study for robust quantum circuits,''
  in \emph{Proceedings of the 41st {IEEE/ACM} International Conference on
  Computer-Aided Design, {ICCAD} 2022, San Diego, California, USA, 30 October
  2022 - 3 November 2022}, T.~Mitra, E.~F.~Y. Young, and J.~Xiong, Eds.\hskip
  1em plus 0.5em minus 0.4em\relax {ACM}, 2022, p. 136. [Online]. Available:
  \url{https://doi.org/10.1145/3508352.3561118}
\BIBentrySTDinterwordspacing

\bibitem{9283531}
C.~J. Wood, ``Special session: Noise characterization and error mitigation in
  near-term quantum computers,'' in \emph{2020 IEEE 38th International
  Conference on Computer Design (ICCD)}, 2020, pp. 13--16.

\bibitem{PhysRevLett.112.050502dd}
\BIBentryALTinterwordspacing
J.~Zhang, A.~M. Souza, F.~D. Brandao, and D.~Suter, ``Protected quantum
  computing: Interleaving gate operations with dynamical decoupling
  sequences,'' \emph{Phys. Rev. Lett.}, vol. 112, p. 050502, Feb 2014.
  [Online]. Available:
  \url{https://link.aps.org/doi/10.1103/PhysRevLett.112.050502}
\BIBentrySTDinterwordspacing

\end{thebibliography}
%%%%%%%%%%%%%%%%%%%%%%%%%%%%%%%%%%%%

\end{document}